\PassOptionsToPackage{hyphens}{url}
\PassOptionsToPackage{table}{xcolor}
\documentclass[conference]{IEEEtran}

\pdfoutput=1

\usepackage{amsmath,balance,multirow,multicol,tabularx}
\usepackage{amssymb}
\usepackage[inline]{enumitem}
\usepackage[binary-units]{siunitx}
\DeclareSIUnit{\nothing}{\relax}
\newcommand{\revise}[1]{{\color{black}#1}}

\usepackage{amsthm}
\newtheorem{theorem}{Theorem}[section]
\newtheorem{proposition}[theorem]{Proposition}
\newtheorem{lemma}[theorem]{Lemma}
\theoremstyle{definition}

\newtheorem{example}[theorem]{Example}

\usepackage[binary-units]{siunitx}
\DeclareSIUnit{\txn}{txn}
\DeclareSIUnit{\batch}{batch}
\newcommand{\Arch}{\mathcal{A}}
\newcommand{\Replicas}{\mathcal{R}}
\newcommand{\Clients}{\mathcal{C}}
\newcommand{\Executors}{\mathcal{E}}
\newcommand{\Storage}{\mathcal{S}}
\newcommand{\Verifier}{\mathcal{V}}
\newcommand{\Client}[1]{\textsc{c}_{#1}}
\newcommand{\Replica}[1]{\textsc{r}_{#1}}
\newcommand{\Executor}[1]{\textsc{e}_{#1}}
\newcommand{\Primary}{\textsc{p}}

\newcommand{\Transaction}{\mathbb{T}}

\newcommand{\Timer}[1]{\tau_{#1}}
\newcommand{\Retimer}[1]{\Upsilon_{#1}}

\newcommand{\ID}[1]{\operatorname{id}(#1)}
\newcommand{\n}[1]{\mathbf{n}_{#1}}
\newcommand{\nn}{\mathbf{n}}

\newcommand{\f}[1]{\mathbf{f}_{#1}}
\newcommand{\nf}[1]{\mathbf{nf}_{#1}}
\newcommand{\kmax}{k_{\text{max}}}

\newcommand{\rw}{\mathbf{rw}}
\newcommand{\e}{\mathbf{e}}
\newcommand{\Vset}{\mathbf{V}}

\newcommand{\pim}{\pi}

\newcommand{\Message}[2]{{#1}(#2)}
\newcommand{\SignMessage}[2]{\langle#1\rangle_{#2}}

\newcommand{\Hash}[1]{H(#1)}
\newcommand{\Digest}{\Delta}

\newcommand{\Name}[1]{\textnormal{\textsc{#1}}}
\newcommand{\BName}[1]{\textnormal{\bf \textsc{#1}}}
\newcommand{\PoE}{\Name{PoE}}
\newcommand{\BFT}{\Name{Bft}}
\newcommand{\CFT}{\Name{Cft}}
\newcommand{\PBFT}{\Name{Pbft}}
\newcommand{\PBFTB}{\BName{Pbft}}

\newcommand{\SBFT}{\Name{Sbft}}

\newcommand{\ResilientDB}{\Name{Resilient\-DB}}
\newcommand{\Serverless}{\Name{ServerlessBFT}}

\newcommand{\ServEB}{\BName{ServBFT-8}}
\newcommand{\ServE}{\Name{ServBFT-8}}
\newcommand{\ServTB}{\BName{ServBFT-32}}
\newcommand{\ServT}{\Name{ServBFT-32}}
\newcommand{\ServerlessCFT}{\Name{ServerlessCFT}}
\newcommand{\ServerlessCFTB}{\BName{ServerlessCFT}}
\newcommand{\NoShim}{\Name{NoShim}}
\newcommand{\NoShimB}{\BName{NoShim}}

\newcommand{\abs}[1]{\lvert #1 \rvert}

\DeclareSIUnit\k{k}
\DeclareSIUnit\ms{ms}
\DeclareSIUnit\GB{GiB}
\DeclareSIUnit\B{B}

\usepackage{fontawesome}
\newcommand{\db}[3][purple]{\node[scale=2,color=black!25!#1] at (#2,#3) {\faDatabase};}
\newcommand{\cog}[2]{\node[scale=2,color=black!75] at (#1 + 0.35,#2 + 0.35) {\faCog};}
\newcommand{\dbcog}[3][purple]{\db[#1]{#2}{#3}\cog{#2}{#3}}
\newcommand{\dbuser}[3][blue]{\node[scale=2,color=black!25!#1] at (#2,#3) {\faUser};}
\newcommand{\dblook}[3][yellow]{\node[scale=2,color=black!25!#1] at (#2,#3) {\faSearch};}

\newcommand{\dbcloud}[3][white]{\node[scale=4.5,color=black!15!#1] at (#2,#3) {\faCloud};}
\newcommand{\dbplane}[3][white]{\node[scale=1.5,color=black!15!#1] at (#2,#3) {\faPaperPlane};}
\newcommand{\dbbigcloud}[3][white]{\node[scale=9,color=black!15!#1] at (#2,#3) {\faCloud};}
\newcommand{\RN}[1]{%
  \textup{\expandafter{\romannumeral#1}}%
}
\usepackage{xcolor}
\usepackage{tikz,pgfplots,pgfplotstable}
\usetikzlibrary{arrows.meta,decorations.pathmorphing,decorations.pathreplacing,patterns}
\tikzset{
    >=Stealth,
    dot/.style={circle,scale=0.35,draw=black,fill=black},
    label/.append style={font=\strut\scriptsize},
    geobftplot/.append style={baseline,scale=0.5425}
}

\usepackage[noend]{algorithmic}
\newenvironment{myprotocol}{
    \hrule
    \smallskip
    \scriptsize
    \algsetup{linenosize=\tiny}
    \begin{algorithmic}[1]
        \newcommand{\SPACE}{\item[]}	
	\newcommand{\GETS}{:=}
        \newcommand{\TITLE}[2]{\item[] \textbf{\underline{##1}} (##2) \textbf{:}\\[0.5pt]}
        \makeatletter
            \newcommand{\EVENT}[1]{\STATE \textbf{event} ##1 \textbf{do}\begin{ALC@g}}
            \newcommand{\ENDEVENT}{\end{ALC@g}}
        \makeatother

	\makeatletter
            \newcommand{\FUNC}[2]{\STATE \textbf{function} \textbf{##1} (##2) \begin{ALC@g}}
            \newcommand{\ENDFUNC}{\end{ALC@g}}
        \makeatother
	\newcommand{\INITIAL}[2]{\item[] \textbf{\underline{##1}} ##2\\[0.5pt]}
	\newcommand{\MC}[1]{{\color{orange}// ##1}}
}{
    \end{algorithmic}
    \smallskip
    \hrule
}

\usepackage{xcolor}
\usepackage{tikz,pgfplots,pgfplotstable}
\usetikzlibrary{arrows.meta}
\tikzset{
    >=Stealth,
    plot/.append style={baseline,scale=0.45},
    label/.append style={font=\strut\footnotesize},
    dot/.style={circle,scale=0.35,draw=black,fill=black},
}
\pgfplotscreateplotcyclelist{mycyclelist}{
    red   ,every mark/.append style={solid,fill=\pgfplotsmarklistfill},mark=*\\
    violet     ,every mark/.append style={solid,fill=\pgfplotsmarklistfill},mark=square*\\
    blue    ,every mark/.append style={solid,fill=\pgfplotsmarklistfill},mark=triangle*\\
    black    ,every mark/.append style={solid,fill=\pgfplotsmarklistfill},mark=pentagon*\\
    brown   ,every mark/.append style={solid,fill=\pgfplotsmarklistfill},mark=halfsquare*\\
    orange  ,every mark/.append style={solid,fill=\pgfplotsmarklistfill},mark=halfcircle*\\
    teal  ,every mark/.append style={solid,fill=\pgfplotsmarklistfill,rotate=180},mark=halfdiamond*\\
    green  ,every mark/.append style={solid,fill=\pgfplotsmarklistfill,rotate=180},mark=diamond*\\
}
\pgfplotscreateplotcyclelist{mycyclelistex}{
    black   ,every mark/.append style={solid,fill=\pgfplotsmarklistfill},mark=*\\
    red     ,every mark/.append style={solid,fill=\pgfplotsmarklistfill},mark=square*\\
}
\pgfplotsset{
    tick label style={font=\Large},
    legend style={font=\normalsize,cells={anchor=west}},
    title style={font=\Large},
    label style={font=\normalsize},
    width=262.5pt,
    height=185pt,
    every axis/.append style={
            ylabel near ticks,
            xlabel near ticks,
            mark size=2.5pt,
            cycle list name=mycyclelist,
            font=\Large,
            y tick label style={
                    /pgf/number format/precision=1,
                    /pgf/number format/fixed,
                    /pgf/number format/fixed zerofill
                }
        },
}

\pgfplotstableread{
    Replicas	BFTServerless   BFTServerlessErrors	PBFT    PBFTErrors	CFTServerless   CFTServerlessErrors	NoShim  NoShimErrors
    4   	    167115	        7743	            173612	10217       207130	        14370	            250000  12023  
    8   	    145350	        4362	            160080	7358        200343	        13645	            250000  12023
    16  	    89829	        6865	            115115	5273        132030	        9292	            250000  12023
    32  	    52745	        2122	            62064 	2397        66864	        3980	            250000  12023
    64  	    25637	        1639	            29385 	1413        29101	        1383	            250000  12023
    128 	    13093	        693	                15388 	1200        15203	        726	                250000  12023
}\graphAtp

\pgfplotstableread{
    Replicas	BFTServerless   BFTServerlessErrors	PBFT    PBFTErrors	CFTServerless   CFTServerlessErrors	NoShim  NoShimErrors
    4		    0.350           0.022             	0.336   0.014       0.33	        0.025             	0.15    0.02
    8		    0.400           0.021             	0.36    0.025       0.343	        0.012             	0.15    0.02
    16		    0.530           0.019             	0.413   0.026       0.353	        0.018             	0.15    0.02
    32		    0.750           0.058             	0.652   0.052       0.588	        0.023             	0.15    0.02
    64		    0.920           0.063             	0.849   0.032       0.814	        0.047             	0.15    0.02
    128		    1.54            0.051             	1.34    0.104       1.288	        0.081             	0.15    0.02
}\graphAlt      

\pgfplotstableread{
    Executors  Serverless8  Serverless8Errors Serverless32  Serverless32Errors
    3	       145350       11230             56940         1711
    5	       124509       7092              52745         3562
    11	       87851        3161              42051         2130
    15	       55832        3912              38442         2925
    21	       47714        3700              38318         2395
}\graphBtp
\pgfplotstableread{
    Executors  Serverless8  Serverless8Errors Serverless32  Serverless32Errors
    3	       0.403        0.019             0.7           0.047
    5	       0.467        0.023             0.75          0.051
    11	       0.678        0.042             0.94          0.05
    15	       0.902        0.032             1.05          0.043
    21	       1.035        0.067             1.05          0.068
}\graphBlt

\pgfplotstableread{
    BatchSize  Serverless8  Serverless8Errors Serverless32  Serverless32Errors
    100  	   20739        1415              6467          1229
    300 	   145350       7644              56940         5705
    500 	   171387       7739              91266         6133
    800    	   204170       6395              105141        7922
    1200       239569       14399             110780        7059
    1600       197049       12648             98941         6982
}\graphCtp
\pgfplotstableread{
    BatchSize  Serverless8  Serverless8Errors Serverless32  Serverless32Errors
    100 	   0.356        0.022             0.61          0.04
    300 	   0.403        0.017             0.696         0.04
    500 	   0.745        0.047             0.872         0.051
    800        0.946        0.04              1.287         0.085
    1200       1.307        0.052             2.351         0.136
    1600       1.523        0.05              2.421         0.091
}\graphClt

\pgfplotstableread{
    Sleep  Serverless8  Serverless8Errors Serverless32  Serverless32Errors
    0  	   145350       8202            56940         4643
    1 	   118907       6221            55822         4725
    2 	   84578        3000            49301         3663
    4      47515        2552            35974         2438
    8      37343        2885            27631         2421
}\graphDtp
\pgfplotstableread{
    Sleep  Serverless8  Serverless8Errors Serverless32  Serverless32Errors
    0 	   0.403        0.013             0.696         0.054
    1 	   1.358        0.077             1.452         0.109
    2 	   2.391        0.118             2.489         0.153
    4      4.772        0.257             5.302         0.164
    8      8.49         0.582             9.524         0.573
}\graphDlt

\pgfplotstableread{
    Latency8  Latency32 Throughput8    Throughput32
    0.281	0.336   11909   7427
    0.286	0.349   25907   21196
    0.297	0.474   53965   45734
    0.300   0.548   103025  57546
    0.314   0.694   127592  56515
    0.326   0.847   142433  56638
    0.362   1.035   146553  53781
    0.409   1.185   153378  53963
    0.478   1.283   158020  56701
    0.523   1.42    159440  56916
    0.555   1.569   159924  56368
}\graphEdata

\pgfplotstableread{
    Regions  Serverless8  Serverless8Errors Serverless32  Serverless32Errors
    5  	     87318        5216              43155         3279
    7 	     87554        2630              44598         3506
    9 	     87685        3705              43609         1518
    11       87760        4868              44092         3085
}\graphFtp
\pgfplotstableread{
    Regions  Serverless8  Serverless8Errors Serverless32  Serverless32Errors
    5 	     0.473        0.023             0.933         0.043
    7 	     0.470        0.036             0.905         0.048
    9 	     0.543        0.019             0.931         0.031
    11       0.530        0.016             0.899         0.05
}\graphFlt

\pgfplotstableread{
    Cores  Serverless8  Serverless8Errors Serverless32  Serverless32Errors
    2 	   24156        1841              10731         1921
    4	   58732        4465              22456         3725
    8	   71731        3571              24637         4238
    12     140752       7254              45752         6817
    16     145350       7564              56940         9372
}\graphGtp
\pgfplotstableread{
    Cores  Serverless8  Serverless8Errors Serverless32  Serverless32Errors
    2	   1.299        0.063             1.923         0.11
    4	   1.08         0.052             1.669         0.075
    8	   0.865        0.028             1.621         0.121
    12     0.563        0.045             1.064         0.044
    16     0.403        0.025             0.696         0.039
}\graphGlt

\pgfplotstableread{
    Conflict  Serverless8  Serverless8Errors Serverless32  Serverless32Errors
    0	      145350       9116              56940         6701
    10	      141841       5553              50144         4838
    20	      127691       4472              45882         4017
    30        111573       7146              39675         4548
    40        91872        6500              35439         3304
    50        83121        5902              30543         2608
}\graphHtp
\pgfplotstableread{
    Conflict  Serverless8  Serverless8Errors Serverless32  Serverless32Errors
    0	      0.41         0.013             0.696         0.048
    10	      0.373        0.011             0.729         0.045
    20	      0.41         0.019             0.714         0.044
    30        0.378        0.03              0.714         0.03
    40        0.378        0.012             0.704         0.055
    50        0.41         0.031             0.695         0.038
}\graphHlt

\pgfplotstableread{
    Sleep  	Serverless   	ServerlessE    PBFT1   PBFT1E   PBFT8    PBFT8E  PBFT16  PBFT16E
    0         	51786     	8542           53682   8743     53682    8743    53682   8743
    300	      	51481     	9924           2000    12       16000    78      32000   1042    
    600 	51472     	7361           1000    15       8000     52      16000   447    
    1200 	51902     	8431           200     12       1600     16      3200    288    
    1700 	51915     	4327           100     10       800      8       1600    80     
    2200 	52272     	3134           66      3        528      5       1056    10     
    2700      	50709     	3515           50      2        400      8       800     16     
}\graphItp
\pgfplotstableread{
    Sleep     Serverless   ServerlessE    PBFT1   PBFT1E   PBFT8   PBFT8E  PBFT16  PBFT16E 
    0	      .76          0.02           .72     0.03     .72     0.03   .72     0.03           
    300	      .88          0.07           2.05    0.12     2.05    0.12   2.05    0.12    
    600 	  .92          0.08           2.16    0.19     2.16    0.19   2.16    0.19    
    1200 	  .97          0.05           2.23    0.22     2.23    0.22   2.23    0.22    
    1700 	  1.51         0.13           2.41    0.02     2.41    0.02   2.41    0.02    
    2200 	  2.02         0.04           2.45    0.14     2.45    0.14   2.45    0.14    
    2700      2.84         0.11           2.64    0.21     2.64    0.21   2.64    0.21    
}\graphIlt

\pgfplotstableread{
    Sleep     Serverless       PBFT1            PBFT8            PBFT16 
    0         0.134098877      0.1293626252     0.1293626252     0.1293626252 
    300	      0.1380433479     3.472222222      0.4340277778     0.2170138889  
    600 	  0.1412169343     6.944444444      0.8680555556     0.4340277778 
    1200 	  0.1652991685     34.72222222      4.340277778      2.170138889
    1700 	  0.196765664      69.44444444      8.680555556      4.340277778
    2200 	  0.2273520899     105.2188552      13.1523569       6.576178451
    2700      0.2629469807     138.8888889      17.36111111      8.680555556
}\graphJcost

\newcommand{\axisNodes}{Number of Replicas ($\nn$)}
\newcommand{\axisExec}{Number of Executors}
\newcommand{\axisSleep}{Execution Length (seconds)}
\newcommand{\axisSleepm}{Increase in Execution Time (milliseconds)}
\newcommand{\axisRegions}{Number of Regions}
\newcommand{\axisCores}{Number of Cores}
\newcommand{\axisConflict}{Conflict Percentage}
\newcommand{\axisTP}{Throughput (\si{\txn\per\second})}
\newcommand{\axisLat}{Latency (\si{\second})}
\newcommand{\axisbatches}{Batch size}

\newcommand{\graphATP}{
    \begin{tikzpicture}[plot]
        \begin{axis}[
                scaled y ticks = false,
                xlabel={\axisNodes},
                ylabel={\axisTP},
                ymin=0,
                ytick={0,40000,80000,120000,160000,200000,250000},
                yticklabels={0,40K,80K,120K,160K,200K,250K},
                xtick={4,8,16,32,64,128},
                xmode=log,
                log ticks with fixed point,
                legend to name={mainlegend2},
                legend columns=-1,
            ]
            \addlegendimage{black,every mark/.append style={solid,fill=\pgfplotsmarklistfill},mark=pentagon*}
            \addlegendentry{\Serverless{}};
            
            \addlegendimage{orange,every mark/.append style={solid,fill=\pgfplotsmarklistfill},mark=halfcircle*}
            \addlegendentry{\ServerlessCFT{}};
            
            \addlegendimage{teal,every mark/.append style={solid,fill=\pgfplotsmarklistfill},mark=triangle*}
            \addlegendentry{\PBFT{}};
            \addlegendimage{green,every mark/.append style={solid,fill=\pgfplotsmarklistfill},mark=triangle*}
            \addlegendentry{\NoShim{}};

            \addplot [black,every mark/.append style={solid,fill=\pgfplotsmarklistfill},mark=pentagon*,error bars/.cd,y dir=both,y explicit] 
                table[x={Replicas},y={BFTServerless},y error={BFTServerlessErrors}] {\graphAtp};
            \addplot [orange,every mark/.append style={solid,fill=\pgfplotsmarklistfill},mark=halfcircle*,error bars/.cd,y dir=both,y explicit] 
                table[x={Replicas},y={CFTServerless},y error={CFTServerlessErrors}] {\graphAtp};
            \addplot [teal,every mark/.append style={solid,fill=\pgfplotsmarklistfill},mark=triangle*,error bars/.cd,y dir=both,y explicit] 
                table[x={Replicas},y={PBFT},y error={PBFTErrors}] {\graphAtp};
            \addplot [green,every mark/.append style={solid,fill=\pgfplotsmarklistfill},mark=triangle*,error bars/.cd,y dir=both,y explicit] 
                table[x={Replicas},y={NoShim},y error={NoShimErrors}] {\graphAtp};
        \end{axis}
    \end{tikzpicture}
}

\newcommand{\graphALat}{
    \begin{tikzpicture}[plot]
        \begin{axis}[
                xlabel={\axisNodes},
                ymin=0,
                ylabel={\axisLat},
                xtick={4,8,16,32,64,128},
                xmode=log,
                log ticks with fixed point,
            ]
            \addplot [black,every mark/.append style={solid,fill=\pgfplotsmarklistfill},mark=pentagon*,error bars/.cd,y dir=both,y explicit] 
                table[x={Replicas},y={BFTServerless},y error={BFTServerlessErrors}] {\graphAlt};
            \addplot [orange,every mark/.append style={solid,fill=\pgfplotsmarklistfill},mark=halfcircle*,error bars/.cd,y dir=both,y explicit] 
                table[x={Replicas},y={CFTServerless},y error={CFTServerlessErrors}] {\graphAlt};
            \addplot [teal,every mark/.append style={solid,fill=\pgfplotsmarklistfill},mark=triangle*,error bars/.cd,y dir=both,y explicit] 
                table[x={Replicas},y={PBFT},y error={PBFTErrors}] {\graphAlt};
            \addplot [green,every mark/.append style={solid,fill=\pgfplotsmarklistfill},mark=triangle*,error bars/.cd,y dir=both,y explicit] 
                table[x={Replicas},y={NoShim},y error={NoShimErrors}] {\graphAlt};
        \end{axis}
    \end{tikzpicture}
}

\newcommand{\graphBTP}{
    \begin{tikzpicture}[plot]
        \begin{axis}[
                title={(\RN{1}) Scaling Executors (Throughput)},
                scaled y ticks = false,
                xlabel={\axisExec},
                ylabel={\axisTP},
                ymin=0,
                ytick={0,30000,60000,90000,120000,150000},
                yticklabels={0,30K,60K,90K,120K, 150k},
                xtick={3,5,11,15,21},
                legend to name={mainlegend},
                legend columns=-1,
            ]

            \addlegendimage{red,every mark/.append style={solid,fill=\pgfplotsmarklistfill},mark=*}
            \addlegendentry{\ServE{}};
            
            \addlegendimage{violet,every mark/.append style={solid,fill=\pgfplotsmarklistfill},mark=square*}
            \addlegendentry{\ServT{}};

            \addplot[red,every mark/.append style={solid,fill=\pgfplotsmarklistfill},mark=*,error bars/.cd,y dir=both,y explicit] 
                table[x={Executors},y={Serverless8},y error={Serverless8Errors}] {\graphBtp};
            \addplot[violet,every mark/.append style={solid,fill=\pgfplotsmarklistfill},mark=square*,error bars/.cd,y dir=both,y explicit] 
                table[x={Executors},y={Serverless32},y error={Serverless32Errors}] {\graphBtp};

        \end{axis}
    \end{tikzpicture}
}

\newcommand{\graphBLat}{
    \begin{tikzpicture}[plot]
        \begin{axis}[
                xlabel={\axisExec},
                ymin=0,
                ylabel={\axisLat},
                xtick={3,5,11,15,21},
                title={(\RN{2}) Scaling Executors (Latency)},
            ]
            \addplot[red,every mark/.append style={solid,fill=\pgfplotsmarklistfill},mark=*,error bars/.cd,y dir=both,y explicit] 
            table[x={Executors},y={Serverless8},y error={Serverless8Errors}] {\graphBlt};
            \addplot[violet,every mark/.append style={solid,fill=\pgfplotsmarklistfill},mark=square*,error bars/.cd,y dir=both,y explicit] 
            table[x={Executors},y={Serverless32},y error={Serverless32Errors}] {\graphBlt};
        \end{axis}
    \end{tikzpicture}
}

\newcommand{\graphCTP}{
    \begin{tikzpicture}[plot]
        \begin{axis}[
                title={(\RN{3}) Batching (Throughput)},
                scaled y ticks = false,
                xlabel={\axisbatches},
                ylabel={\axisTP},
                ymin=0,
                ytick={0,40000,80000,120000,160000,200000,240000},
                yticklabels={0,40K,80K,120K,160K,200K,240K},
                xtick={100,300,500,800,1200,1600},
                xticklabels={10,100,200,1000,5000,8000},
            ]
            \addplot[red,every mark/.append style={solid,fill=\pgfplotsmarklistfill},mark=*,error bars/.cd,y dir=both,y explicit]  table[x={BatchSize},y={Serverless8},y error={Serverless8Errors}] {\graphCtp};
            \addplot[violet,every mark/.append style={solid,fill=\pgfplotsmarklistfill},mark=square*,error bars/.cd,y dir=both,y explicit]  table[x={BatchSize},y={Serverless32},y error={Serverless32Errors}] {\graphCtp};
        \end{axis}
    \end{tikzpicture}
}

\newcommand{\graphCLat}{
    \begin{tikzpicture}[plot]
        \begin{axis}[
                xlabel={\axisbatches},
                ymin=0,
                ylabel={\axisLat},
                xtick={100,300,500,800,1200,1600},
                xticklabels={10,100,200,1000,5000,8000,},
                title={(\RN{4}) Batching (Latency)},
            ]
            \addplot[red,every mark/.append style={solid,fill=\pgfplotsmarklistfill},mark=*,error bars/.cd,y dir=both,y explicit]  table[x={BatchSize},y={Serverless8},y error={Serverless8Errors}] {\graphClt};
            \addplot[violet,every mark/.append style={solid,fill=\pgfplotsmarklistfill},mark=square*,error bars/.cd,y dir=both,y explicit]  table[x={BatchSize},y={Serverless32},y error={Serverless32Errors}] {\graphClt};
        \end{axis}
    \end{tikzpicture}
}

\newcommand{\graphDTP}{
    \begin{tikzpicture}[plot]
        \begin{axis}[
                title={(\RN{5}) Expensive Execution (Throughput)},
                scaled y ticks = false,
                xlabel={\axisSleep},
                ylabel={\axisTP},
                ymin=0,
                ytick={0,30000,60000,90000,120000,150000},
                yticklabels={0,30K,60K,90K,120K, 150k},
                xtick={0,1,2,4,8},
            ]
            \addplot[red,every mark/.append style={solid,fill=\pgfplotsmarklistfill},mark=*,error bars/.cd,y dir=both,y explicit]  table[x={Sleep},y={Serverless8},y error={Serverless8Errors}] {\graphDtp};
            \addplot[violet,every mark/.append style={solid,fill=\pgfplotsmarklistfill},mark=square*,error bars/.cd,y dir=both,y explicit]  table[x={Sleep},y={Serverless32},y error={Serverless32Errors}] {\graphDtp};
        \end{axis}
    \end{tikzpicture}
}

\newcommand{\graphDLat}{
    \begin{tikzpicture}[plot]
        \begin{axis}[
                xlabel={\axisSleep},
                ymin=0,
                ylabel={\axisLat},
                xtick={0,1,2,4,8},
                title={(\RN{6}) Expensive Execution (Latency)},
            ]
            \addplot[red,every mark/.append style={solid,fill=\pgfplotsmarklistfill},mark=*,error bars/.cd,y dir=both,y explicit]  table[x={Sleep},y={Serverless8},y error={Serverless8Errors}] {\graphDlt};
            \addplot[violet,every mark/.append style={solid,fill=\pgfplotsmarklistfill},mark=square*,error bars/.cd,y dir=both,y explicit]  table[x={Sleep},y={Serverless32},y error={Serverless32Errors}] {\graphDlt};
        \end{axis}
    \end{tikzpicture}
}

\newcommand{\graphFTP}{
    \begin{tikzpicture}[plot]
        \begin{axis}[
                title={(\RN{7}) Executor Distribution  (Throughput)},
                scaled y ticks = false,
                xlabel={\axisRegions},
                ylabel={\axisTP},
                ymin=20000,
                ytick={0,20000,40000,60000,80000,100000},
                yticklabels={0,20K,40K,60K,80K, 100k},
                xtick={5,7,9,11,13},
            ]
            \addplot[red,every mark/.append style={solid,fill=\pgfplotsmarklistfill},mark=*,error bars/.cd,y dir=both,y explicit]  table[x={Regions},y={Serverless8},y error={Serverless8Errors}] {\graphFtp};
            \addplot[violet,every mark/.append style={solid,fill=\pgfplotsmarklistfill},mark=square*,error bars/.cd,y dir=both,y explicit]  table[x={Regions},y={Serverless32},y error={Serverless32Errors}] {\graphFtp};
        \end{axis}
    \end{tikzpicture}
}

\newcommand{\graphFLat}{
    \begin{tikzpicture}[plot]
        \begin{axis}[
                xlabel={\axisRegions},
                ymin=.3,
                ylabel={\axisLat},
                xtick={5,7,9,11,13},
                title={(\RN{8}) Executor Distribution (Latency)},
            ]
            \addplot[red,every mark/.append style={solid,fill=\pgfplotsmarklistfill},mark=*,error bars/.cd,y dir=both,y explicit]  table[x={Regions},y={Serverless8},y error={Serverless8Errors}] {\graphFlt};
            \addplot[violet,every mark/.append style={solid,fill=\pgfplotsmarklistfill},mark=square*,error bars/.cd,y dir=both,y explicit]  table[x={Regions},y={Serverless32},y error={Serverless32Errors}] {\graphFlt};
        \end{axis}
    \end{tikzpicture}
}

\newcommand{\graphGTP}{
    \begin{tikzpicture}[plot]
        \begin{axis}[
                title={(\RN{9}) Computing Power (Throughput)},
                scaled y ticks = false,
                xlabel={\axisCores},
                ylabel={\axisTP},
                ymin=0,
                ytick={0,30000,60000,90000,120000,150000},
                yticklabels={0,30K,60K,90K,120K, 150k},
                xtick={2,4,8,12,16},
            ]
            \addplot[red,every mark/.append style={solid,fill=\pgfplotsmarklistfill},mark=*,error bars/.cd,y dir=both,y explicit]  table[x={Cores},y={Serverless8},y error={Serverless8Errors}] {\graphGtp};
            \addplot[violet,every mark/.append style={solid,fill=\pgfplotsmarklistfill},mark=square*,error bars/.cd,y dir=both,y explicit]  table[x={Cores},y={Serverless32},y error={Serverless32Errors}] {\graphGtp};
        \end{axis}
    \end{tikzpicture}
}

\newcommand{\graphGLat}{
    \begin{tikzpicture}[plot]
        \begin{axis}[
                xlabel={\axisCores},
                ymin=0,
                ylabel={\axisLat},
                xtick={2,4,8,12,16},
                title={(\RN{10}) Computing Power (Latency)},
            ]
            \addplot[red,every mark/.append style={solid,fill=\pgfplotsmarklistfill},mark=*,error bars/.cd,y dir=both,y explicit]  table[x={Cores},y={Serverless8},y error={Serverless8Errors}] {\graphGlt};
            \addplot[violet,every mark/.append style={solid,fill=\pgfplotsmarklistfill},mark=square*,error bars/.cd,y dir=both,y explicit]  table[x={Cores},y={Serverless32},y error={Serverless32Errors}] {\graphGlt};
        \end{axis}
    \end{tikzpicture}
}

\newcommand{\graphHTP}{
    \begin{tikzpicture}[plot]
        \begin{axis}[
                title={(\RN{11}) Conflicting Transactions (Throughput)},
                scaled y ticks = false,
                xlabel={\axisConflict},
                ylabel={\axisTP},
                ymin=0,
                ytick={0,30000,60000,90000,120000,150000},
                yticklabels={0,30K,60K,90K,120K, 150k},
                xtick={0,10,20,30,40,50},
            ]
            \addplot[red,every mark/.append style={solid,fill=\pgfplotsmarklistfill},mark=*,error bars/.cd,y dir=both,y explicit]  table[x={Conflict},y={Serverless8},y error={Serverless8Errors}] {\graphHtp};
            \addplot[violet,every mark/.append style={solid,fill=\pgfplotsmarklistfill},mark=square*,error bars/.cd,y dir=both,y explicit]  table[x={Conflict},y={Serverless32},y error={Serverless32Errors}] {\graphHtp};
        \end{axis}
    \end{tikzpicture}
}

\newcommand{\graphHLat}{
    \begin{tikzpicture}[plot]
        \begin{axis}[
                xlabel={\axisConflict},
                ymin=0,
                ylabel={\axisLat},
                xtick={0,10,20,30,40,50},
                title={(\RN{12}) Conflicting Transactions (Latency)},
            ]
            \addplot[red,every mark/.append style={solid,fill=\pgfplotsmarklistfill},mark=*,error bars/.cd,y dir=both,y explicit]  table[x={Conflict},y={Serverless8},y error={Serverless8Errors}] {\graphHlt};
            \addplot[violet,every mark/.append style={solid,fill=\pgfplotsmarklistfill},mark=square*,error bars/.cd,y dir=both,y explicit]  table[x={Conflict},y={Serverless32},y error={Serverless32Errors}] {\graphHlt};
        \end{axis}
    \end{tikzpicture}
}

\newcommand{\graphE}{
    \begin{tikzpicture}[plot,scale=1]
        \begin{axis}[
                width=362.5pt,
                scaled y ticks = false,
                xlabel={\axisLat},
                ylabel={\axisTP},
                ymin=0,
                ytick={0,40000,80000,120000,160000},
                yticklabels={0,40K,80K,120K,160K},
            ]
            \addplot table[x={Latency8},y={Throughput8}] {\graphEdata};
            \addplot table[x={Latency32},y={Throughput32}] {\graphEdata};
            \legend{\ServE{},\ServT{}}
        \end{axis}
    \end{tikzpicture}
}

\newcommand{\graphITP}{
    \begin{tikzpicture}[plot]
        \begin{axis}[
                width=270pt,
                scaled y ticks = false,
                xlabel={\axisSleepm},
                ylabel={\axisTP},
                ymin=0,
                ymode=log,
                log basis y={2},
                ytick={100,1000,10000,50000},
                yticklabels={100,1K,10K,50K},
                xtick={0,300,600,1200,1700,2200,2700},
                xticklabels={0,50,100,500,1000,1500,2000},
		legend to name={mainlegend3},
                legend columns=-1,
            ]
	    	\addlegendimage{red,every mark/.append style={solid,fill=\pgfplotsmarklistfill},mark=*}
            	\addlegendentry{\ServT{}};

	    	\addlegendimage{violet,every mark/.append style={solid,fill=\pgfplotsmarklistfill},mark=square*}
            	\addlegendentry{\PBFT{}-1-ET}

		\addlegendimage{violet,every mark/.append,mark=triangle*}
            	\addlegendentry{\PBFT{}-8-ET}

		\addlegendimage{violet,every mark/.append,mark=pentagon*}
            	\addlegendentry{\PBFT{}-16-ET}

            \addplot[red,every mark/.append style={solid,fill=\pgfplotsmarklistfill},mark=*,error bars/.cd,y dir=both,y explicit]  table[x={Sleep},y={Serverless},y error={ServerlessE}] {\graphItp};
            \addplot[violet,every mark/.append style={solid,fill=\pgfplotsmarklistfill},mark=square*,error bars/.cd,y dir=both,y explicit]  table[x={Sleep},y={PBFT1},y error={PBFT1E}] {\graphItp};
            \addplot[violet,every mark/.append, dashed,mark=triangle*,error bars/.cd,y dir=both,y explicit]  table[x={Sleep},y={PBFT8},y error={PBFT8E}] {\graphItp};
            \addplot[violet,every mark/.append, densely dashdotted,mark=pentagon*,error bars/.cd,y dir=both,y explicit]  table[x={Sleep},y={PBFT16},y error={PBFT16E}] {\graphItp};
        \end{axis}
    \end{tikzpicture}
}

\newcommand{\graphJ}{
    \begin{tikzpicture}[plot]
        \begin{axis}[
                width=270pt,
                scaled y ticks = false,
                xlabel={\axisSleepm},
                ylabel={Cost (\si{cents\per\k\txn})},
                ymin=0,
                ymode=log,
                log basis y={2},
                ytick={0.01,1,10,100},
                yticklabels={0,1,10,100},
                xtick={0,300,600,1200,1700,2200,2700},
                xticklabels={0,50,100,500,1000,1500,2000},
            ]
            \addplot[red,every mark/.append style={solid,fill=\pgfplotsmarklistfill},mark=*,error bars/.cd,y dir=both,y explicit]  table[x={Sleep},y={Serverless}] {\graphJcost};
            \addplot[violet,every mark/.append style={solid,fill=\pgfplotsmarklistfill},mark=square*,error bars/.cd,y dir=both,y explicit]  table[x={Sleep},y={PBFT1}] {\graphJcost};
            \addplot[violet,every mark/.append, dashed,mark=triangle*,error bars/.cd,y dir=both,y explicit]  table[x={Sleep},y={PBFT8}] {\graphJcost};
            \addplot[violet,every mark/.append, densely dashdotted,mark=pentagon*,error bars/.cd,y dir=both,y explicit]  table[x={Sleep},y={PBFT16}] {\graphJcost};
        \end{axis}
    \end{tikzpicture}
}

\begin{document}
\title{Reliable Transactions in Serverless-Edge Architecture}

%% The "author" command and its associated commands are used to define the authors and their affiliations.
\author{\IEEEauthorblockN{Suyash Gupta$^{*}$,\quad Sajjad Rahnama,\quad Erik Linsenmayer,\quad Faisal Nawab$^{\dagger}$,\quad Mohammad Sadoghi}
\IEEEauthorblockA{%
    Exploratory Systems Lab\\ 
    University of California, Davis\\ 
   $^{\dagger}$University of California, Irvine\\
   $^{*}$University of California, Berkeley
}}

\maketitle

\begin{abstract}
\revise{
Modern edge applications demand novel solutions 
where edge applications do not have to rely on a single cloud provider 
(which cannot be in the vicinity of every edge device)
or dedicated edge servers (which cannot scale as clouds) 
for processing compute-intensive tasks.
A recent computing philosophy, Sky computing, proposes giving 
each user ability to select between available cloud providers.

In this paper, we present our serverless-edge co-design, which extends the 
Sky computing vision.
In our serverless-edge co-design, we expect edge devices to collaborate and 
spawn required number of serverless functions.
This raises several key challenges: 
(1) how will this collaboration take place, 
(2) what if some edge devices are compromised, and 
(3) what if a selected cloud provider is malicious. 
Hence, we design \Serverless{}, 
the first protocol to guarantee Byzantine fault-tolerant (\BFT{}) transactional 
flow between edge devices and serverless functions.
We present an exhaustive list of attacks and their solutions on our serverless-edge co-design.
Further, we extensively benchmark our architecture on a variety of parameters.
}

\end{abstract}


\section{Introduction}
\label{s:intro}
\revise{
This paper introduces \Serverless{}, the first protocol to guarantee Byzantine fault-tolerant (\BFT{}) transactional 
flow between edge devices and serverless functions.
The design of \Serverless{} is motivated from the recent introduction of {\em Sky Computing}, 
which envisages utility computing in a multi-cloud environment~\cite{cloud-to-sky,sky-above-clouds}.
Sky computing propounds the design of an {\em inter-cloud broker} that takes as input 
a client program and output specifications and selects the best cloud providers to execute 
the client program.
Such a broker is extremely desirable for the edge and Internet of Things (IoT) applications, which run 
on edge devices, such as smart devices, sensors, UAVs, and phones, 
that have limited compute power and memory~\cite{cisco-report}.

On the one hand, existing edge applications expect response latency 
in the order of tens of milliseconds~\cite{edgechain,edge-tail-latency,edge-iot}. 
On the other hand, they are forced to delegate compute-intensive tasks to 
a specific third-party cloud provider such as AWS and Azure~\cite{homecloud,cloudlets}.
A recent way to solve this dilemma is to install dedicated {\em edge-servers} 
that are closer to the edge devices~\cite{ekya,wattedge}.
}
These edge servers are installed and maintained by the enterprise behind the application~\cite{assaia,wind-iot}.
If any hardware crashes, then the enterprise may need to purchase new hardware.
\revise{
Moreover, with ever-growing application needs, these servers are impossible to seamless scale as third-party clouds.
}

\revise{
\Serverless{} realizes the Sky computing vision in edge computing by giving the edge applications flexibility to
select any of the available cloud providers.
As a result, the edge application can select different cloud providers based on the location of its users.%
\footnote{\revise{
At present, switching cloud providers is common for most applications due to geo-political reasons and government regulations~\cite{one-trust}.
}}
However, moving data across cloud providers degrades system performance and is expensive.
So, we take a step further and permit edge applications to make use of serverless technology, which 
(i) decouples storage, compute, and network, 
(ii) supports pay-as-you-go model where the enterprise pays only for the resources used, and 
(iii) supports auto-scaling policies~\cite{aft,serverless-trends}.
We refer to this interaction as {\em serverless-edge co-design} 
as it promotes light-weight tasks at the edge while compute-intensive tasks are done at the serverless cloud.
Our serverless-edge co-design targets low latency by allowing edge devices to spawn serverless functions at the 
nearest cloud.

Our serverless-edge co-design also presents several research challenges, which we enlist next.

(1) {\em Task distribution between edge and serverless.}
Our \Serverless{} protocol requires an edge application to push its compute-intensive task to the cloud 
by spawning serverless functions (for simplicity, we refer to these functions as {\em executors}). 
To do so, we need to design a {\em compatibility layer}. 
We build this compatibility layer on top of edge devices and refer to it as a {\em shim}.
At shim, the edge devices collaborate and spawn serverless executors for executing compute-intensive client requests.

(2) {\em Lack of Trust at Shim.}
As edge devices may belong to different parties, which may not trust each other, 
it is hard for these devices to collaborate. 
Hence, our \Serverless{} protocol runs a traditional \BFT{} protocol 
to allow these edge devices reach a consensus~\cite{pbft,geobft}. 
This consensus decides which edge device will spawn the desired number of executors and 
the order in which client requests are processed.
For consensus, we opt for \BFT{} protocols as they are resilient to malicious attacks.
Further, depending on the location and nature of edge devices, 
\Serverless{} permits various shim designs:
} 
a single shim of all devices running \PBFT~\cite{pbft} consensus, 
multiple dependent shims of devices spread globally, running GeoBFT~\cite{geobft}, and
multiple independent shims running \Name{Ahl}~\cite{ahl}, Sharper~\cite{sharper}, or RingBFT~\cite{ringbft}.
For simplicity, in this paper, we assume a single shim of $3\f{}+1$ devices where up to $\f{}$ devices can act malicious.

\revise{
(3) {\em Lack of Trust at Serverless cloud.}
Depending on the application requirement, shim may spawn serverless executors at one or more available 
cloud providers in the vicinity. 
Hence, there is again a lack of trust: some cloud providers may have mal-intent or may have poor QoS (crashed or failed executors)~\cite{upright,cores-dont-count}.
}
As a result, \Serverless{} requires the shim to spawn $2\f{}+1$ executors and permits up to $\f{}$ of them to fail.
This extra spawning is not new; Yahoo's Hadoop also executes the same 
code multiple times to reduce latency due to stragglers~\cite{hadoop}.

\revise{
(4) {\em Private Data access and retrieval.}
Recent reports illustrate that around $90\%$ of the industries are not only sticking with 
their existing on-premise servers, but also scaling them up~\cite{on-premise-growth,on-premise-business}.
For at least $65\%$ of these industries, the key reason for maintaing on-premise servers 
is to protect their consumer data from data-breaches and attacks~\cite{on-premise-growth}.
In our serverless-edge co-design, we adhere to this design choice and 
assume that all the client data is stored in an on-premise storage at the enterprise. 
As a result, the enterprise can control access to the data.
Hence, edge devices or executors lack rights to update the storage, but may request read access to the same.
For updates to the storage, 
we write a lightweight wrapper ({\em verifier}) around the storage that
collects execution results, updates the data-store, and 
forwards the results to the clients.
}

\revise{
Furthermore, we observe several other new challenges with our architecture:}
(i) Byzantine shim devices may spawn less executors, for which we need to hold them accountable.
(ii) During execution, executors may need to read data from the storage.
(iii) If the client transactions are conflicting and their read-write sets are unknown until execution,
we may have to abort such transactions.

\revise{
We envision our serverless-edge architecture to seamlessly integrate with existing edge applications.
}
To realize this goal experimentally, we design a shim of nodes and 
require them to spawn AWS Lambda functions as executors.
On each shim node, we install \ResilientDB's light-weight and multi-threaded consensus framework ~\cite{geobft,rcc,r-evalpaper,blockchain-book,flexi-trust}.
We evaluate our \Serverless{} protocol on {\em eight} distinct parameters. 
Our results illustrate that \Serverless{} can facilitate shims of up to $128$ devices in $11$ global regions. 
Further, in our experiments, we are easily able to spawn $21$ executors in parallel (could not scale further due to limits by cloud provider),
and the peak throughput achieved by our \Serverless{} protocol is \SI{240}{k} txns/s while the minimum latency incurred is \SI{30}{ms}.

In this paper, we make the following {\em contributions}:

\begin{itemize}[wide,nosep]
\revise{
\item We design of a novel serverless-edge co-design that meets the vision of Sky computing and 
helps design low latency reliable edge applications where edge devices can select cloud providers 
based on desired output specifications.
}

\item In our serverless-edge architecture, we neither trust the edge devices nor the serverless executors. 
Hence, we introduce a novel protocol \Serverless{} that manages the flow of a client request in our 
serverless-edge architecture and shields the system against arbitrary results and malicious attacks.

\item We enlist possible attacks in our serverless-edge architecture and present 
solutions to recover the system.

\item Our \Serverless{} protocol presents algorithms to handle conflicting transactions 
with or without the knowledge of read-write available to shim nodes prior to execution.

\end{itemize}

\section{Motivation and Use Case}
\label{s:moti}

\begin{figure}[t]
    \centering
    \begin{tikzpicture}[xscale=0.47,yscale=0.85]
	\dbplane[orange]{0}{-2}
	\dbplane[orange]{4}{-2}
	\dbplane[orange]{0}{-3}
	\dbplane[orange]{4}{-3}
	\dbuser{1.9}{-4.5}

	\path (-1, -1.65) edge[-] (4.8, -1.65);
	\path (-1, -1.65) edge[-] (-1, -3.45);
	\path (-1, -3.45) edge[-] (4.8, -3.45);
	\path (4.8, -1.65) edge[-] (4.8, -3.45);

	\path (0.9, -2) edge[<->] (3.1, -2);
	\path (0.9, -2) edge[<->] (0.9, -3);
	\path (3.1, -2) edge[<->] (3.1, -3);
	\path (0.9, -3) edge[<->] (3.1, -3);
	\path (0.9, -2) edge[<->] (3.1, -3);
	\path (3.1, -2) edge[<->] (0.9, -3);

	%% Executors
	\dbcloud[white]{0}{0}
	\cog{-1.1}{-0.6}
	\cog{0.3}{-0.6}
	\dbcloud[white]{3.5}{0}
	\cog{3}{-0.6}

	\dblook[yellow]{1}{2}
	\dbcog[brown]{3}{2}

        \node[above=5pt,label] at (1.9, -5.6)  {\scriptsize Client};
	\node[above=5pt,label] at (2, -3.8)  {\scriptsize Shim};
        \node[below=5pt,label] at (3.8, -0.2)  {\scriptsize Executors};
	\node[above=5pt,label] at (-0.5, 1.5)  {\scriptsize Verifier};
	\node[above=5pt,label] at (4.8, 1.5)  {\scriptsize Storage};

	\path (1.95, -4.2) edge[->] node[blue,pos=-0.3,left=4.9pt,label] {\scriptsize (1) Client txn} (1.95, -3.5);
	
	\node[above=1pt,label,blue] at (4.5, -4)    {\scriptsize (2) Consensus};
	\node[above=1pt,label,blue] at (4.7, -4.25) {\scriptsize between};
	\node[above=1pt,label,blue] at (5.0, -4.5)  {\scriptsize edge nodes};

	\path (1.5, -1.6) edge[->] node[blue,below=-5pt,label] {\scriptsize (3) Spawn executors} (0.2, -0.5);
	\path (1.5, -1.6) edge[->] (3, -0.5);

	\draw[->] (0.4, 0.5) -- (1.4, 1.5) node[blue,pos=0.85,left=1pt,label] {\scriptsize (5) Send Results};
	\draw[->] (2.5, 0.5) -- (1.4, 1.5);

	\draw[->] (0.4, 0.5) -- (3, 1.5) node[blue,pos=0.8,right=4.5pt,label] {\scriptsize (4) Fetch data};
	\path (2.5, 0.5) edge[->] (3, 1.5);

	\path (1.5, 2) edge[<->] (2.2, 2);

	%%% Second Design
	\dbplane[teal]{9}{-2}
	\dbplane[teal]{13}{-2}
	\dbplane[teal]{10}{-3}
	\dbplane[teal]{12}{-3}
	\dbuser{11}{-4.5}

	\dbbigcloud[white]{11}{1}
	\dbcog[brown]{10}{1.5}
	\cog{10}{0}
	\cog{12}{0}

	\path (10.95, -4.2) edge[->] node[blue,pos=-0.3,left=2.3pt,label] {\scriptsize (1) Client txn} (10.95, -3.5);
	\node[blue,label] at (13.5,2.2) {\scriptsize (2) Collect and Process};

	\draw[->] (9, -1.7) -- (9, -0.3);
	\draw[->] (13, -1.7) -- (13, -0.3);
	\draw[->] (10, -2.7) -- (10, -0.3);
	\draw[->] (12, -2.7) -- (12, -0.3);
	\draw[-] (7, 2.5) -- (7, -5);

	\node[above=5pt,label] at (11, -5.6)  {\scriptsize Client};
	\node[below=5pt,label] at (11, 0.4)  {\scriptsize Edge Servers};
	\node[above=5pt,label] at (11, -6)  {\scriptsize (b)};
	\node[above=5pt,label] at (1.9, -6) {\scriptsize (a)};

    \end{tikzpicture}
    \caption{Illustration of (a) Serverless-Edge architecture employing the \Serverless{} protocol and (b) architecture prevalent in existing edge applications.}
     \label{fig:serverless-arch}
\end{figure}
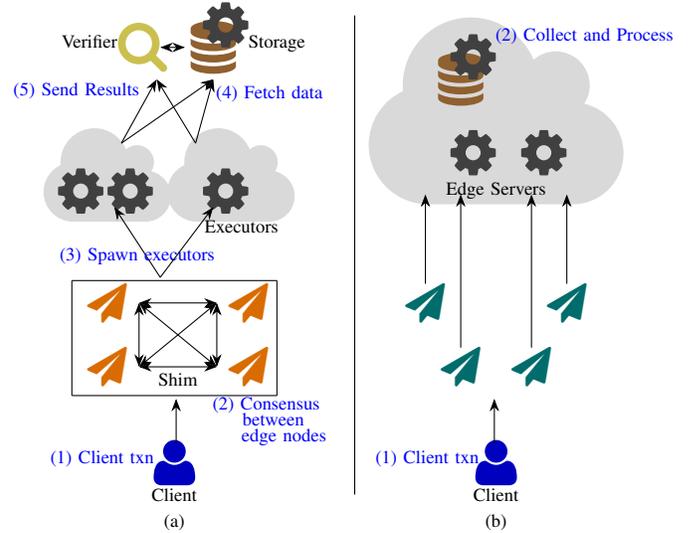

\revise{
The motivations behind our serverless-edge co-design are the emerging use cases of edge-computing, such as AR/VR video-streaming and  
Unmanned Aerial Vehicles (UAVs). 
These applications require massive data-processing as they need to run ML models to train data on-flight or provide the user 
useful insights.
The key challenge these applications face is the rapidly changing user characteristics.

We consider a real-world use case of UAVs as a motivating example for this work~\cite{amazon-uav}.
In recent years, UAVs have been adopted by e-commerce industries, such as Amazon and Walmart, for product deliveries.
These UAVs help to securely and quickly transport user goods in a cost-efficient manner.
During the delivery process, each UAV travels over multiple geographical locations and performs an array of tasks, such as 
navigation, image recognition, and live video-streaming.
  
In Figure~\ref{fig:serverless-arch}(b), we illustrate the traditional way of computing for UAVs, 
where each UAV offloads all the collected data to the dedicated edge servers for processing.
In this model, UAVs are forced to communicate with dedicated servers.
When the server is in the vicinity, the communication round-trip costs are low; 
otherwise, they are high.
Each edge server executes the requests from various UAVs in an ordered-fashion.
}
Moreover, these servers need to be continuously scaled, 
new software needs to be installed, and OS needs to be updated, which makes them a financially expensive choice.

\revise{
In Figure~\ref{fig:serverless-arch}(a), we reimagine the UAV delivery operation in our serverless-edge co-design.
Switching to our serverless-edge model, allows UAVs in the vicinity to interact with each other and 
act as a shim that spawns serverless executors to process collected data. 
To alleviate concerns regarding round-trip costs, 
the shim is permitted to opt for services from local cloud providers.
}
In fact, \revise{shim can} spawn executors at multiple clouds and wait for whichever responds earliest.
\revise{
As there is a lack of trust among shim devices and executors, we have our \Serverless{} protocol 
to manage all the transactional flow in a byzantine fault-tolerant manner.
}

\revise{
{\em Byzantine failures in the wild.} 
Do real-world systems face more than just crash failures? 
Unfortunately, yes.
}
Existing systems suffer from {\em omission} failures where nodes can crash~\cite{spanner}, and 
{\em arbitrary} failures where nodes can act in an unexpected manner~\cite{upright}.
Almost all real-world applications handle omission failures using protocols based on 
Paxos-family~\cite{paxos,raft}.
However, the true challenge is to bulwark the system system against often overlooked arbitrary failures:
Google's UpRight~\cite{upright} provides fault-tolerance against byzantine failures, 
Google has also observed corrupt execution errors~\cite{cores-dont-count}, and
Cloudflare observed a misbehaving switch sending incorrect messages~\cite{cloudflare}.
Hence, it is better to guard system against these failures.

\section{Preliminaries}
\label{s:prelim}
We make standard assumptions as made by any \BFT{} system~\cite{pbft,geobft,sharper,rcc}.
We represent our serverless-edge architecture $\Arch$ through a quintuple, 
$\Arch = \{\Clients, \Replicas, \Executors, \Storage, \Verifier \}$, where
we use $\Clients$ to denote the set of clients, 
$\Replicas$ to denote the shim of edge devices or nodes, 
$\Executors$ to denote the serverless executors, 
$\Verifier$ and $\Storage$ to denote the verifier and data-store.
\revise{
As described in Section~\ref{s:intro}, we assume an on-premise data-store maintained by the enterprise, 
while the verifier is a lightweight wrapper around the data-store. 
Hence, both verifier and storage are assumed to be honest and trusted.
}

{\em Fault-Tolerance Requirement at Shim.}
We use the notation $\n{\Replicas} = \abs{\Replicas}$ to represent total number of edge nodes in $\Arch$.
At most $\f{\Replicas}$ of these nodes are byzantine and can crash-fail or act arbitrarily; 
$\n{\Replicas} \ge 3\f{\Replicas} + 1$.
The remaining $2\f{\Replicas} + 1$ nodes are honest and follow the protocol.

{\em Authenticated Communication.}
To exchange messages among different components, we employ 
Digital Signatures (\Name{DS}) and Message Authentication Codes (\Name{MAC})~\cite{cryptobook}.
To represent a message $m$ signed by a component $\Replica{}$ using \Name{DS}, we use the notation 
{\bf $\SignMessage{m}{\Replica{}}$}.
Anyone who has the signer's public-key can verify this signature.
\revise{
One of the common ways to exchange public-keys is through a public-key certificates~\cite{publickey-management}.
}
For \Name{MAC}s, signer and verifier use a common key, which is kept secret.
\revise{
We use Diffie-Hellman key exchange for securely sharing secret keys.
}
 In rest of the text, any message $m$ that does not indicate the identity of the signer 
implies the use of \Name{MAC}.
Although \Name{MAC}s offer higher throughput than \Name{DS}, 
\Name{DS} guarantee {\em non-repudiation}~\cite{pbft,bc-processing}.
We also employ a \emph{collision-resistant hash function} $\Hash{\cdot}$ 
to map a value $v$ to a constant-sized digest $\Hash{v}$. 
We use a function $\ID{}$ to assign an identifier to each node $\Replica{} \in \Replicas$ and 
each executor $\Executor{} \in \Executors{}$.
We assume that byzantine components can neither impersonate honest
components, nor subvert cryptographic constructs.
We {\em do not} make any assumptions on the behavior of the clients.
We term a message as {\em well-formed} if it passes all the 
cryptographic and other necessary checks.

\subsection{Serverless Cloud Assumptions}
\label{ss:serverless-assume}
We expect access to one or more serverless clouds such as AWS Lambda and Google Functions.
\revise{
These serverless cloud should permit edge nodes to seamlessly upload the desirable code or transactions for 
processing as per the application specifications.
}
For simplicity, in rest of the text, we assume that the shim nodes access only one cloud provider for
{\em spawning executors} to execute client transactions.
However, there is {\em no free food} as these serverless clouds follow a {\em pay-per-use} model 
where whoever spawns executors also pays for their use~\cite{berkeley-serverless}.
We expect these clouds to meet the following:
\begin{itemize}[wide]
\item {\em Fault-Tolerance: }
To handle arbitrary faults at the serverless cloud, 
we spawn $\n{\Executors} \ge 2\f{\Executors}+1$ executors, and 
assume that at most $\f{\Executors}$ are byzantine.
Prior works have shown that $2\f{\Executors}+1$ executors guarantee successful execution of a 
transaction in the byzantine setting~\cite{separation-execution}.
This leads us to observe the following:
\begin{enumerate}[wide,nosep]
\item The values for $\f{\Executors}$ and $\f{\Replicas}$ may or may not be same.

\item In Section~\ref{s:conflicts}, we illustrate that if the transactions are conflicting, 
then we need an additional $\f{\Executors}$ executors to prevent an indistinguishable byzantine attack.
\end{enumerate}

\item {\em Identity: }
We expect each spawned executor to be assigned a unique pair of public-private key, which
it uses to digitally sign a message.

\item {\em Accountability: }
Each executor is spawned by some shim node that pays for this service. 
Hence, we expect that no executor can spawn more executors.
Further, the expected number of executors {\em to be spawned} by shim nodes  
is known to all the components of our architecture.

\item {\em Payment.} 
As executors are spawned by shim nodes,
it implies that the spawner will be billed by the cloud provider. 
Hence, post successful consensus of a transaction, the edge application's
enterprise pays the spawner a fixed amount to cover its expenses.

\end{itemize}

\section{Architecture}

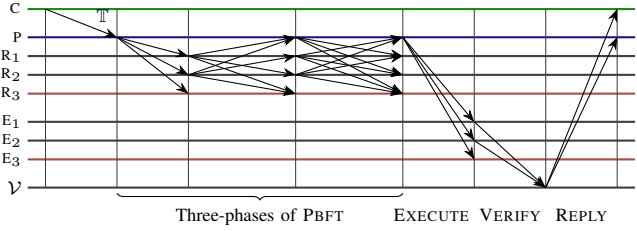
\begin{figure}[t]
    \centering
    \begin{tikzpicture}[yscale=0.25,xscale=0.95]
        \draw[thick,draw=black!75] (0.75, 11.5) edge[green!50!black!90] ++(8.5, 0)
                                   (0.75,   7) edge[red!50!black!70] ++(8.5, 0)
                                   (0.75,   8) edge ++(8.5, 0)
                                   (0.75,   9) edge ++(8.5, 0)
                                   (0.75,   10) edge[blue!50!black!90] ++(8.5, 0)
					
				   (0.75,   5.5) edge ++(8.5, 0)
                                   (0.75,   4.5) edge ++(8.5, 0)
                                   (0.75,   3.5) edge [red!50!black!70]++(8.5, 0)

				   (0.75,   2) edge ++(8.5, 0);

        \draw[thin,draw=black!75] (1, 2) edge ++(0, 9.5)
                                  (2, 2) edge ++(0, 9.5)
				  (3, 2) edge ++(0, 9.5)
                                  (4.5, 2) edge ++(0, 9.5)
                                  (6, 2) edge ++(0, 9.5)
				  (7, 2) edge ++(0, 9.5)
				  (8,  2) edge ++(0, 9.5)
				  (9,2) edge ++(0, 9.5);

        \node[left] at (0.8, 7) {\scriptsize $\Replica{3}$};
        \node[left] at (0.8, 8) {\scriptsize $\Replica{2}$};
        \node[left] at (0.8, 9) {\scriptsize $\Replica{1}$};
        \node[left] at (0.8, 10) {\scriptsize $\Primary$};
        \node[left] at (0.8, 11.5) {\scriptsize $\Client{}$};

	\node[left] at (0.8, 5.5) {\scriptsize $\Executor{1}$};
        \node[left] at (0.8, 4.5) {\scriptsize $\Executor{2}$};
        \node[left] at (0.8, 3.5) {\scriptsize $\Executor{3}$};

	\node[left] at (0.8, 2) {\scriptsize $\Verifier{}$};

        \path[->] (1, 11.5) edge node[above,pos=0.8] {\scriptsize $\Transaction$} (2, 10)
                  (2, 10) edge (3, 9)
                          edge (3, 8)
                          edge (3, 7)
                           
                  (3, 9) edge (4.5, 10)
			 edge (4.5, 8)
			 edge (4.5, 7)

		  (3, 8) edge (4.5, 10)
			 edge (4.5, 9)
			 edge (4.5, 7)

		  (4.5, 10) edge (6, 9)
			 edge (6, 8)
			 edge (6, 7)

		  (4.5, 9) edge (6, 10)
			 edge (6, 8)
			 edge (6, 7)

		  (4.5, 8) edge (6, 10)
			 edge (6, 9)
			 edge (6, 7)

                  (6, 10) edge (7, 5.5)
                  (6, 10) edge (7, 4.5)
                  (6, 10) edge (7, 3.5)

		  (7, 5.5) edge (8, 2)
                  (7, 4.5) edge (8, 2)

		  (8, 2) edge (9, 11.5)
	  	  (8, 2) edge (9, 10)	  
                           ;

	\draw [decorate, decoration = {brace}] (6,1.7) --  (2,1.7);

        \node[below] at (4, 1.5) {\scriptsize \strut Three-phases of \PBFT{}};

        \node[below] at (6.4, 1.5) {\scriptsize \strut\Name{Execute}};
	\node[below] at (7.5, 1.5) {\scriptsize \strut\Name{Verify}};
	\node[below] at (8.5, 1.5) {\scriptsize \strut\Name{Reply}};

    \end{tikzpicture}
    \caption{Schematic representation of the transactional flow in \Serverless{} protocol. 
	Given a client transaction $\Transaction$, the nodes of the shim work together to order this transaction, 
	following which the primary $\Primary{}$ invokes the executors at the serverless cloud to execute $\Transaction$.
	Post execution, the executors send their results to the verifier, which replies to the client.
	}
    \label{fig:serverless-protocol}
\end{figure}

\revise{
We now discuss in detail the \BFT{} transactional flow guaranteed by our \Serverless{} 
protocol in the serverless-edge co-design.
In Figure~\ref{fig:serverless-protocol}, we schematically present this flow;
}
the shim consists of $\n{\Replicas}=4$ edge nodes and $\n{\Executors}=3$ executors are spawned per transaction.
\revise{
For understandability, we will periodically refer to the UAV use case of Section~\ref{s:moti}.
}

As stated earlier, shim can have different abstractions and can run any \BFT{} protocol.
In this paper, we assume a single shim of $3\f{\Replicas}+1$ 
and require shim nodes to run the \PBFT{}~\cite{pbft} protocol.
\PBFT{} is considered as a representative \BFT{} protocol as all the other protocols follow its design.
\PBFT{} protocol works in {\em views}. 
For each view, one node is designated as the {\em primary} and is responsible for successful 
completion of consensuses in that view.
If the primary acts malicious, the view is changed and the primary is replaced.

\subsection{Client Request and Response}
Any user that accesses the edge application becomes a client in our system.
\revise{
E.g., each UAV that requires data-processing from the cloud acts as a client 
and packages its request as a transaction.
}
A client $\Client{}$ send a message $\SignMessage{\Transaction}{\Client{}}$
to the primary node $\Primary{}$%
\footnote{
Some \BFT{} protocols require a client request to be sent to all the nodes.
}
of the current view $v$ of the shim when it wants to process a transaction $\Transaction$.
Notice that $\Client{}$ employs \Name{DS} to sign this message (refer to Figure~\ref{alg:serverless}, Line~\ref{alg:client-send}).
The client $\Client{}$ marks $\SignMessage{\Transaction}{\Client{}}$ as processed when 
it receives a $\Name{Response}$ message from the verifier $\Verifier{}$.
As $\Client{}$ knows that $\Verifier{}$ is a trusted entity in our infrastructure, 
it readily accepts the response (Line~\ref{alg:serv:res}).

\subsection{Shim Ordering}
\revise{
\Serverless{} assigns each shim node (e.g. UAV) an identifier, $0,1,2,...,\n{\Replicas{}}$. 
Initially, the shim node with identifier $0$ is designated as the primary $\Primary{}$ of the shim.
}
On receiving a client request $\SignMessage{\Transaction}{\Client{}}$, 
$\Primary{}$ checks if $\SignMessage{\Transaction}{\Client{}}$ is well-formed. 
If this is the case, $\Primary$ initiates the \PBFT{} protocol as follows.

\begin{itemize}[wide]
\item {\em Pre-prepare.} 
The primary $\Primary{}$ assigns a sequence number $k$ to the well-formed client message $m := \SignMessage{\Transaction}{\Client{}}$ 
and sends it as a $\Name{Preprepare}$ message to all the nodes of the shim.
This $\Name{Preprepare}$ message also includes a digest $\Digest = \Hash{m}$, which is used in future 
communication to save space.
Notice that the primary signs this message using \Name{MAC}, which provide sufficient guarantees for this phase. 
When a node $\Replica{} \in \Replicas{}$  receives a $\Name{Preprepare}$ message from the primary $\Primary{}$ of view $v$,
it runs the message through a series of checks.
If the checks are successful, then $\Replica{}$ agrees to support the order $k$ for this client request by
broadcasting a $\Name{Prepare}$ message.

\item {\em Prepare.}
When a node $\Replica{}$ receives identical $\Name{Prepare}$ messages from $2\f{\Replicas{}}+1$ distinct nodes 
(can include its own message to reach the count), 
it marks the request $m$ as {\em prepared} and broadcasts a $\Name{Commit}$ message.
We require each node $\Replica{}$ to use $\Name{DS}$ to sign the $\Name{Commit}$ message.

\item {\em Commit.}
When $\Replica{}$ receives identical $\Name{Commit}$ messages from $2\f{\Replicas{}}+1$ nodes,
it marks $m$ as {\em committed}.

\end{itemize}

{\bf \em Remark.}
\PBFT{} requires two phases of quadratic communication complexity. 
\revise{
Instead, shim can employ \BFT{} protocols like \PoE~\cite{poe} and \SBFT~\cite{sbft} 
that guarantee linear communication with the help of advanced cryptographic schemes like threshold signatures.
Note: in our architecture, the edge devices are acting as both clients and shim nodes. 
}

\subsection{Serverless Optimistic Execution}
Once $\Primary{}$ commits a request, 
\Serverless{} requires $\Primary$ to connect with the serverless cloud
and spawn $\n{\Executors{}}$ executors. 
\Primary{} sends each of these executors an $\Name{Execute}$ message (Line~\ref{alg:spawn}),
which includes a {\em certificate} $\mathfrak{C}$; 
a set of signatures of $2\f{\Replicas{}}+1$ distinct shim nodes and
proves that these  nodes agreed to order this request (Line~\ref{alg:serv:ds-commit}).
Prior to executing the transaction $\Transaction{}$, 
each executor $\Executor{} \in \Executors{}$ checks if the certificate $\mathfrak{C}$ is valid.

During execution, $\Executor{}$ may need to access the value of read-write sets ($\rw$). 
Hence, it connects with the storage $\Storage{}$ and {\em fetches} the required data (Lines~\ref{alg:rw-check}-\ref{alg:rw-fetch}). 
However, executors {\em do not write} to the storage. 
Any intermediate results are stored locally.
Further, these executors do not communicate with each other. 
Post execution, each executor $\Executor{}$ sends a $\Name{Verify}$ message to the verifier $\Verifier{}$, 
which includes the computed result $r$, 
certificate $\mathfrak{C}$, and accessed read-write sets $\rw$.

{\bf \em Remark.} 
\revise{
We allow shim to spawn either stateless or stateful executors~\cite{berkeley-serverless,serverless-computing}.
Stateful executors have memory and remember the results of last execution. 
By definition, severless executors are ``fleeting'' and return after execution; a common way to assign these 
executors memory is by having a layer that stores computed results~\cite{aft}.
To employ stateful executors in our model, we would need \BFT{} guarantees on the additional layer.
Hence, we focus on stateless executors.
}
Including $\mathfrak{C}$ in the \Name{Execute} and $\Name{Verify}$ messages helps to detect byzantine 
attacks (\S~\ref{ss:verifier-abort}).
\revise{
Further, by employing threshold signatures,
we can reduce the size of the certificate. 
Threshold signatures allow combining $2\f{\Replicas{}}+1$ signatures into 
a single signature.
}

\subsection{Verifier and Concurrency Control}
The verifier $\Verifier{}$ is a lightweight wrapper around the data-store $\Storage$ 
and is assumed to be {\em correct and trusted}.
The verifier collects well-formed $\Name{Verify}$ messages from the executors (in set $\Vset$) and 
once it has a quorum of matching results that do not violate the {\em concurrency control constraints}, 
it updates the data-store. 
It performs these tasks in the following {\em order}:
\begin{enumerate}[nosep,wide]
\item If set $\Vset$ has at least $\f{\Executors}+1$ matching $\Name{Verify}$ messages, $\Verifier$
marks the transaction as {\em matched}. 
Following this, $\Verifier{}$ ignores any other $\Name{Verify}$ message for 
$\SignMessage{\Transaction}{\Client{}}$ (Line~\ref{alg:f1}).

\item If $k$ is the sequence number for $\SignMessage{\Transaction}{\Client{}}$ and 
$\kmax$ is the sequence number of last {\em validated} request, then if 
$\kmax \neq k$, $\Verifier{}$ places the $k$-th request in the list $\pim$ (Line~\ref{alg:store}).

\item If $\kmax = k$, $\Verifier{}$ checks if the value of the read-write sets $\rw$ of the $k$-th request
is same as that in the data-store $\Storage$ (Lines~\ref{alg:read}-\ref{alg:check}).
If the {\em read sets match}, $\Verifier{}$ sends the client and the shim primary $\Name{Response}$ messages
and updates the write sets at the storage in accordance with the result $r$ 
(Lines~\ref{alg:reply}-\ref{alg:update}).
\revise{
Note: matching read-write sets is only required when the transactions are conflicting. 
We discuss this in Section~\ref{s:conflicts}.
}

\item Next, $\Verifier{}$ increments $\kmax$ and checks if $\pim$ includes the transaction 
with sequence number $\kmax$. 
If so, it removes the $\kmax$-th transaction from $\pim$ and 
runs steps in Lines~\ref{alg:loop}-\ref{alg:loop-check}.

\end{enumerate}

\begin{figure}[t]
    \begin{myprotocol}
	\INITIAL{Initialization:}{\newline
	{\color{orange}
	// $\kmax \GETS$ sequence number of the next request to be verified by $\Verifier$.\newline
	// $\pim \GETS \emptyset$ (list of requests marked matched at verifier)
	}}
	\vspace{1mm}
	%\SPACE

	%% Client request
        \TITLE{Client-role}{used by client $\Client{}$ to request transaction $\Transaction{}$}
        \STATE Sends $\SignMessage{\Transaction{}}{\Client{}}$ to the primary $\Primary{}$.\label{alg:client-send}
        \STATE Awaits receipt of message $\Message{\Name{Response}}{\SignMessage{\Transaction{}}{\Client{}}, k, r}$ from 
		$\Verifier{}$.\label{alg:client-wait}
        \STATE Considers $\Transaction{}$ executed, with result $r$, as the $k$-th transaction.\label{alg:serv:res}
        \SPACE

	%% Primary receives client request
        \TITLE{Primary-role}{running at the primary node $\Primary{}$}
	 \EVENT{$\Primary{}$ receives $\SignMessage{\Transaction{}}{\Client{}}$}
		 \STATE Calculate digest $\Digest := \Hash{\SignMessage{\Transaction{}}{\Client{}}}$.
		 \STATE Broadcast $\Message{\Name{Preprepare}}{\SignMessage{\Transaction{}}{\Client{}},\Digest,k}$ to all nodes (order at sequence $k$).
	 \ENDEVENT
	 \SPACE

	 %% Primary receives Commit message
	 \EVENT{$\Primary{}$ receives $\nf{\Replicas{}}$ $m := \SignMessage{\Message{\Name{Commit}}{\Digest,k}}{\Replica{}}$ messages such that:
	       \begin{enumerate}[nosep]
                	\item each message $m$ is well-formed and is sent by a distinct node $\Replica{} \in \Replicas{}$.
         	\end{enumerate}
	 }
	       \STATE $\mathfrak{C}$ := set of \Name{DS} of these $\nf{\Replicas{}}$ messages. \MC{Certificate}\label{alg:serv:ds-commit}
	       \STATE Send $\SignMessage{\Message{\Name{Execute}}{\SignMessage{\Transaction{}}{\Client{}},\mathfrak{C},m,\Delta}}{\Primary{}}$ 
	       	to all executors $\Executor{} \in \Executors{}$. \MC{Serverless access} \label{alg:spawn}
         \ENDEVENT
	 \SPACE

	\TITLE{Non-Primary role}{running at a node $\Replica{} \in \Replicas$}
	 %% Non-Primary receives Pre-Prepare message.
	 \EVENT{$\Replica{}$ receives $\Message{\Name{Preprepare}}{\SignMessage{\Transaction{}}{\Client{}},\Digest,k}$ from $\Primary{}$ such that:
	    \begin{enumerate}[nosep]
		\item message is well-formed, and $\Replica{}$ did not accept a $k$-th proposal from $\Primary{}$.
            \end{enumerate}
	 }
		\STATE Broadcast $\Message{\Name{Prepare}}{\Digest,k}$ to all nodes in $\Replicas{}$.	
	 \ENDEVENT
	 \SPACE

	\TITLE{All nodes role}{running at the node $\Replica{}$}
	%% Replica receives Prepare message.
	 \EVENT{$\Replica{}$ receives $\Message{\Name{Prepare}}{\Digest,k}$ messages from $\nf{\Replicas{}}$ nodes such that:
	    \begin{enumerate}[nosep]
                \item each message is well-formed and is sent by a distinct node, $\Replica{*} \in \Replicas{}$.
            \end{enumerate}
	 }
		\STATE Broadcast $\SignMessage{\Message{\Name{Commit}}{\Digest,k}}{\Replica{}}$ to all nodes in 
				$\Replicas{}$.
	 \ENDEVENT
	 \SPACE

	\TITLE{Executor-role}{running at the executor $\Executor{} \in \Executors$}
	 %% Executor receives Execute message.
	 \EVENT{$\Executor{}$ receives $\SignMessage{\Message{\Name{Execute}}{\SignMessage{\Transaction{}}{\Client{}},\mathfrak{C},m,\Delta}}{\Primary{}}$ from $\Primary{}$ such that:
	    \begin{enumerate}[nosep]
		\item message is well-formed,
		\item $m := \Message{\Name{Commit}}{\Digest,k}$, and
		\item Certificate $\mathfrak{C}$ includes $\nf{\Replica{}}$ distinct \Name{DS} on $m$.
            \end{enumerate}
	 }
		\WHILE{$\Transaction{}$ not executed}
			\STATE $\rw{} :=$ Read-write sets for $\Transaction{}$.
			\IF{Need the current state of $\rw{}$ \MC{Storage access}} \label{alg:rw-check}
				\STATE Fetch $\rw{}$ state (values) from storage $\Storage{}$ \label{alg:rw-fetch}
			\ENDIF
		\ENDWHILE

		\STATE $r :=$ Result of executing $\Transaction{}$

		\STATE Send $\Message{\Name{Verify}}{\SignMessage{\Transaction{}}{\Client{}},\mathfrak{C},m,\rw,r}$ to verifier $\Verifier{}$. \MC{Communication with verifier}
	 \ENDEVENT
	 \SPACE

	\TITLE{Verifier-role}{running at the verifier $\Verifier{}$}
	 %% Non-Primary receives Pre-Prepare message.
	 \EVENT{$\Verifier{}$ receives $m' := \Message{\Name{Verify}}{\SignMessage{\Transaction{}}{\Client{}},A,m,\rw,r}$ message from an executors such that:
	       \begin{enumerate}[nosep]
                	\item $m'$ is well-formed and is sent by a distinct executor $\Executor{} \in \Executors{}$, and
			\item $m := \Message{\Name{Commit}}{\Digest,k}$.
         	\end{enumerate}
	 }
		\STATE Add $m'$ to $\Vset$.\label{alg:vset}
	 \ENDEVENT
	 \SPACE

	 \EVENT{Set $\Vset$ has $\f{\Executors}+1$ identical $m'$ \GETS $\Message{\Name{Verify}}{\SignMessage{\Transaction{}}{\Client{}},A,m,\rw,r}$ messages}\label{alg:f1}
		\IF{$k = \kmax$ \MC{Next request in order.}}\label{alg:kmax}
			\STATE Run function {\bf ccheck($\pim$)} 
			\WHILE{$\kmax$-th transaction is in $\pim$ \MC{Other requests}}  \label{alg:loop}
				\STATE Run function {\bf ccheck($\pim$)} \label{alg:loop-check}
			\ENDWHILE
		\ELSE
			\STATE Store $m'$ in $\pim$.\label{alg:store}
		\ENDIF
	 \ENDEVENT
	 \SPACE

	 \FUNC{ccheck}{list: $\pi$}\label{alg:ccheck}
			
		\STATE $\rw' := $ Current state of $\rw{}$ fetched from storage $\Storage$.\label{alg:read}
		\IF{$\rw' = \rw$ \MC{Concurrency control check}}\label{alg:check}
			\STATE Send $\SignMessage{\Message{\Name{Response}}{\Delta,r}}{\Verifier{}}$ 
	       		to the client $\Client{}$ and primary $\Primary{}$. \MC{Reply to client.}\label{alg:reply}
			\STATE Update corresponding $\rw{}$ with $r$ at the storage $\Storage$.\label{alg:update}
		\ENDIF
		\STATE $\kmax = \kmax + 1$.\label{alg:inc}
		
	 \ENDFUNC
    \end{myprotocol}
    \caption{Byzantine Fault-Tolerant transaction processing by \Serverless{} protocol in the serverless-edge architecture.}
    \label{alg:serverless}
    \vspace{-2mm}
\end{figure}

These concurrency control checks ensure that consistent updates are 
written to the storage.

\subsection{System Guarantees}
\label{ss:guarantees}
We now state the guarantees offered by our different components of our serverless-edge architecture.
\begin{description}[nosep]
\item[\em Shim Consistency.] 
If an honest  node commits a transaction $\Transaction{}$, then all the honest
nodes commit $\Transaction{}$.

\item[\em Shim Non-Divergence.] 
If two honest nodes order a transaction $\Transaction{}$ at sequence number $k$ and $k'$, 
then $k = k'$.

\item[\em Shim Termination.]
If an honest client sends a transaction $\Transaction{}$, then 
an honest node will eventually commit $\Transaction{}$.

\item[\em Executor Termination.]
If an honest  primary sends an $\Name{Execute}$ message for 
transaction $\Transaction{}$, then 
an honest executor will execute $\Transaction{}$.

\item[\em Verifier Non-Divergence.] 
If the shim commits a transaction $\Transaction{}$ at sequence $k$, 
then the verifier will eventually update the corresponding result at the storage at order $k$.

\end{description}

Together, shim consistency, shim non-divergence,  and verifier non-divergence 
imply {\em safety}, 
while shim termination and executor termination imply {\em liveness}. 
Our \Serverless{} protocol guarantees safety in an asynchronous environment where the 
messages can get lost, delayed, or duplicated, and byzantine components can collude or 
act arbitrarily. 
To guarantee liveness, our \Serverless{} protocol expects periods of synchrony.
Note: our \Serverless{} offers standard safety and liveness guarantees, also offered by 
other systems~\cite{pbft,poe,geobft,ahl,sharper}.

\section{Tackling Byzantine Attacks}
\label{s:byz-attacks}
\revise{
In our architecture, at most $\f{\Replicas}$ shim nodes and $\f{\Executors}$ 
serverless executors can act byzantine. 
If the primary of shim is honest, then byzantine nodes cannot affect 
the ongoing transactional flow. 
Similarly, byzantine executors can either provide incorrect result or ignore execution, but 
as there are at least $\f{\Executors}+1$ honest executors, $\Name{Execute}$ messages 
sent by honest primaries will be processed.
Hence, following is an exhaustive list of attacks on our design.
}

\begin{enumerate}[wide,nosep,label=(\roman*)]
\item\label{A:req-ignore} {\em Request Suppression.} 
If the primary of shim is byzantine, it can try to prevent consensus on some client requests. 

\item\label{A:rep-dark} {\em Nodes in Dark.}
If shim's primary is byzantine, it can keep up to $\f{\Replicas{}}$ 
honest shim nodes in {\em dark} by not involving them in consensuses.

\item\label{A:ver-flood} {\em Verifier Flooding.}
Byzantine components can flood the verifier with requests that have been already verified.

\end{enumerate}

Next, we present algorithms to recover from these attacks.

\subsection{Request Suppression}
\label{ss:req-suspension}
In the serverless-edge architecture, byzantine components can work together 
to deny service to one or more clients.
This request suppression attack can take three different forms:

\begin{enumerate}[nosep,wide,label=(\roman*)]
\item \label{sup:ignore}{\em Request Ignorance.}
If the shim's primary node $\Primary{}$ is byzantine, 
it can willfully drop a request $m$ from a client $\Client{}$,  
or indefinitely delay consensus on $m$.

\item \label{sup:unsuccess}{\em Unsuccessful Consensus.}
A byzantine primary $\Primary{}$  
may involve less than $2\f{\Replicas}+1$ nodes in consensus 
on a client request $m$. 
As a result, these nodes will not reach consensus on $m$.

\item \label{sup:less}{\em Less Executors.}
A byzantine primary $\Primary{}$ may permit consensus on 
a client request $m$, but disallow its execution by spawning less 
than $\n{\Executors{}}$ serverless executors.
In such a case, the verifier $\Verifier{}$ will not receive 
$\f{\Executors}+1$ matching execution results.

\end{enumerate}

To detect these attacks, we setup three distinct {\em timers} at various components of our architecture.
\begin{itemize}[nosep,wide]
\item {\em Client timer.} 
Our \Serverless{} protocol requires each client $\Client{}$ to start a timer 
$\Timer{m}$ prior to sending its request $m$ to the primary $\Primary{}$.
When $\Client{}$ receives a $\Name{Response}$ message for $m$ from the verifier $\Verifier$, 
it stops $\Timer{m}$.

\item {\em Node timer.}
Our \Serverless{} protocol requires each node $\Replica{} \in \Replicas{}$ to start a timer 
$\Timer{m}$ when it receives a well-formed $\Name{Preprepare}$ message 
for a client request $m$ from the primary $\Primary{}$.
When $\Replica{}$ marks $m$ as committed, it stops $\Timer{m}$.

\item {\em Node re-transmission timer.}
If a non-primary node $\Replica{} \in \Replicas{}$ receives an $\Name{Error}$ 
message from the verifier $\Verifier{}$ (see Section~\ref{sss:verifier-error}) 
then $\Replica{}$ forwards the $\Name{Error}$ message to the primary $\Primary{}$ and 
starts the re-transmission timer $\Retimer{}$.
When $\Replica{}$ receives a corresponding $\Name{Ack}$ message from $\Verifier{}$, it stops $\Retimer{}$.

\end{itemize}
 
In the case the timers of $\Client{}$ or $\Replica{}$ expire, 
the respective component detects a request suspension attack and 
initiates the following mechanisms for recovery from this attack.

\begin{figure}[t]
    \begin{myprotocol}
	%% Client request
        \TITLE{Client-role}{running at the client $\Client{}$}
	 \EVENT{$\Client{}$'s timer $\Timer{m}$ for request $m := \SignMessage{\Transaction{}}{\Client{}}$ timeouts}
        	\STATE Sends $\SignMessage{\Transaction{}}{\Client{}}$ to the verifier $\Verifier{}$.
		\STATE Restarts $\Timer{m}$. 
        	\IF{Figure~\ref{alg:serverless}, Lines~\ref{alg:client-wait} and~\ref{alg:serv:res} are
			successful \MC{Receives $\f{\Replicas{}} +1$ matching responses}}
			\STATE Cancel $\Timer{m}$
		\ENDIF
	 \ENDEVENT
        \SPACE

	\TITLE{Verifier-role}{running at the verifier $\Verifier{}$}
	 %% Non-Primary receives Pre-Prepare message.
	 \EVENT{$\Verifier{}$ receives a well-formed request $m := \SignMessage{\Transaction{}}{\Client{}}$ from client $\Client{}$}
		\IF{Previously sent $\Name{Response}$ for $m$}
			\STATE Resends message $\SignMessage{\Message{\Name{Response}}{\Delta,r}}{\Verifier{}}$ to $\Client{}$.
		\ELSIF{$m$ exists in list $\pim$ \MC{Waiting for consensus of $\kmax$-th request}}
			\STATE Broadcasts $\SignMessage{\Message{\Name{Error}}{\kmax}}{\Verifier}$ to all shim nodes.
		\ELSIF{Did not receive any $\Name{Verify}$ message for $\SignMessage{\Transaction{}}{\Client{}}$}
			\STATE Broadcasts $\SignMessage{\Message{\Name{Error}}{\SignMessage{\Transaction}{\Client{}}}}{\Verifier}$ to all shim nodes.  \MC{Missing Request}
		\ELSE 
			\STATE Broadcasts $\SignMessage{\Message{\Name{Replace}}{\SignMessage{\Transaction}{\Client{}}}}{\Verifier}$ to all shim nodes.  \MC{Byzantine Primary}
		\ENDIF
         \ENDEVENT
	 \SPACE

	\TITLE{Node-role}{running at the node $\Replica{}$}
	%% Replica receives Prepare message.
	 \EVENT{$\Replica{}$ receives $\SignMessage{\Message{\Name{Error}}{\SignMessage{\Transaction{}}{\Client{}}}}{\Verifier}$ or $\SignMessage{\Message{\Name{Error}}{\kmax}}{\Verifier}$ from $\Verifier{}$}
		\STATE Start a timer $\Retimer{}$.
		\STATE Forward the $\Name{Error}$ message to the primary $\Primary{}$.
	 \ENDEVENT
	 \SPACE

	 \EVENT{$\Replica{}$'s timer $\Timer{m}$ {\bf or} $\Retimer{m}$ timeout {\bf or} $\Replica{}$ receives $\SignMessage{\Message{\Name{Replace}}{\SignMessage{\Transaction}{\Client{}}}}{\Verifier}$ from $\Verifier{}$}
		\STATE Run the {\em view-change} protocol to replace $\Primary{}$
	 \ENDEVENT

    \end{myprotocol}
    \caption{Actions performed by various participants of the serverless-edge infrastructure in response to a request suppression attack.}
    \label{alg:recover-reqsup}
\end{figure}

\subsubsection{Client action on timeout}
If a client $\Client{}$'s timer $\Timer{m}$ timeouts, 
then $\Client{}$ forwards its request to the verifier $\Verifier$ 
and restarts its timer (refer to Figure~\ref{alg:recover-reqsup}). 
In specific, each time $\Client{}$'s timer expires, after some exponential backoff, 
it re-sends its request to $\Verifier{}$ until it receives a $\Name{Response}$ message 
from $\Verifier{}$.

\subsubsection{Verifier action on receiving client request}
\label{sss:verifier-error}
When the verifier $\Verifier{}$ receives a request $m := \SignMessage{\Transaction}{\Client{}}$ from client $\Client{}$, 
it first determines if it has seen $\SignMessage{\Transaction}{\Client{}}$ till now or not. 
If $\Verifier{}$ has not received any $\Name{Verify}$ messages for $\SignMessage{\Transaction}{\Client{}}$,
it sends $\SignMessage{\Message{\Name{Error}}{\SignMessage{\Transaction}{\Client{}}}}{\Verifier}$ message to 
all the nodes in the shim.
Otherwise, there can be only three cases:
\begin{enumerate}[nosep,wide,label=(\roman*)]
\item $\Verifier{}$ did send a $\Name{Response}$ message for $\SignMessage{\Transaction}{\Client{}}$, 
so it simply resends the $\Name{Response}$ message.

\item $\SignMessage{\Transaction}{\Client{}}$ resides in $\pim$.
\revise{
Further, assume that it was ordered by shim at some sequence number $k$. 
So $\kmax < k$, and 
$\Verifier{}$ is waiting for the request with sequence number $\kmax$.
}
Unless the $\kmax$-th request is validated by $\Verifier{}$, succeeding requests 
cannot be processed.
\revise{
So, $\Verifier$ needs to notify shim nodes about the missing request at sequence $\kmax$, 
and it does so by sending $\SignMessage{\Message{\Name{Error}}{\kmax}}{\Verifier}$ 
to all the shim nodes.
Note: this gap between $\kmax$ and $k$ could have been created by byzantine primary.
}

\item $\Verifier{}$ did not receive $\f{\Executors}+1$ matching $\Name{Verify}$ messages 
for $\SignMessage{\Transaction}{\Client{}}$.
This can only occur if the primary is byzantine.
So, $\Verifier{}$ sends $\SignMessage{\Message{\Name{Replace}}{\SignMessage{\Transaction}{\Client{}}}}{\Verifier}$ 
to all the shim nodes.
\end{enumerate}

Once $\Verifier{}$ successfully verifies the request at sequence number $\kmax$ or $\SignMessage{\Transaction}{\Client{}}$, 
$\Verifier{}$ creates a corresponding $\SignMessage{\Message{\Name{Ack}}{\kmax}}{\Verifier}$ 
or $\SignMessage{\Message{\Name{Ack}}{\SignMessage{\Transaction}{\Client{}}}}{\Verifier}$ message and 
broadcasts it to shim.

\subsubsection{Node action on \Name{Error} message}
When a shim node $\Replica{} \in \Replicas{}$ receives an $\Name{Error}$ message 
from the verifier, it can only conclude the following: 
\begin{itemize}[nosep,wide]
\item $\Replica{}$ received $\SignMessage{\Message{\Name{Error}}{\kmax}}{\Verifier}$ message and
has either committed or not committed the request at sequence number $\kmax$.

\item $\Replica{}$ received $\SignMessage{\Message{\Name{Error}}{\SignMessage{\Transaction}{\Client{}}}}{\Verifier}$ message and 
has either committed or not committed the request $\SignMessage{\Transaction}{\Client{}}$.
\end{itemize}

Irrespective of these cases, the node $\Replica{}$ starts a re-transmit timer $\Retimer{}$ to track the 
behavior of the primary.
Next, it forwards the received $\Name{Error}$ message to the primary. 
If the timer $\Retimer{}$ expires before $\Replica{}$ receives a corresponding acknowledgment message
($\SignMessage{\Message{\Name{Ack}}{\kmax}}{\Verifier}$ or 
$\SignMessage{\Message{\Name{Ack}}{\SignMessage{\Transaction}{\Client{}}}}{\Verifier}$)  from the verifier $\Verifier$, 
$\Replica{}$ concludes that the primary is byzantine and requests a view-change.
Hence, the onus is on the primary to guarantee consensus and execution.

\subsubsection{Node action on timeout}
\label{sss:node-timeout}
When the timer $\Timer{m}$ for a node $\Replica{} \in \Replicas{}$ expires, 
$\Replica{}$ concludes that the shim's primary for view $v$ is byzantine, and 
it requests primary replacement by broadcasting a $\Name{ViewChange}$ message.
We employ \PBFT's {\em view-change} protocol to replace a byzantine primary.
A node $\Replica{}$'s request for change of view from $v$ to $v+1$ is only successful 
if it receives support of at least $2\f{\Replicas{}}+1$ nodes, that is, 
at least $2\f{\Replicas{}}+1$ shim nodes must broadcast $\Name{ViewChange}$ messages. 
Replacing the current primary requires designating another shim node as the next primary. 
Like \PBFT{}, we assume nodes have a pre-decided order of becoming the primary.
As a result, when the replica designated as the primary for view $v+1$ receives $\Name{ViewChange}$ requests from at least 
$2\f{\Replicas{}}+1$ nodes, 
it assumes the role of the primary and broadcasts a $\Name{NewView}$ message to 
bring all the nodes to the same state. 
Similarly, when a node $\Replica{}$ receives a $\Name{Replace}$ message from the verifier $\Verifier{}$, 
it initiates the view-change protocol to replace the primary $\Primary{}$ to view $v$.
We defer the details for the exact view-change protocol to the original \PBFT{} paper~\cite{pbft}.

\subsection{Shim Nodes in Dark}
\label{ss:shim-dark}
If the primary $\Primary{}$ is byzantine, it may attempt 
to only include $2\f{\Replicas{}}+1$ nodes in consensus as only $2\f{\Replicas{}}+1$ nodes are needed to mark any request 
as prepared and committed.
As a result, the remaining $\f{\Replicas{}}$ nodes will be in dark.
Next, we explain what we mean by being in dark.
\begin{enumerate}[nosep,wide,label=(\roman*)]
\item \label{dark:exclude}{\em Node Exclusion.}
A byzantine primary $\Primary{}$ can exclude up to $\f{\Replicas}$ honest nodes from 
consensuses by not sending them the $\Name{Preprepare}$ messages for client requests. 

\item \label{dark:equi}{\em Equivocation.}
A byzantine primary $\Primary{}$ can equivocate by associating two 
client requests with the same sequence number $k$. 
If $\Primary{}$ is clever, it will ensure that one of these client requests  
is committed by at least $\f{\Replicas}+1$ honest nodes while the remaining $\f{\Replicas}$ honest nodes do not  
commit any request at sequence number $k$.
\end{enumerate}

The key challenge to resolving the attack~\ref{dark:exclude} is that it is {\em impossible to detect}.
In this attack, the byzantine primary $\Primary{}$ is clever and does not want to risk replacement. 
Hence, $\Primary{}$ facilitates  continuous consensus on incoming client requests 
by at least $\f{\Replicas}+1$ honest nodes.
As a result, the remaining $\f{\Replicas{}}$ nodes are unable to trigger view-change by themselves.

\begin{lemma}
If at most $\f{\Replicas}$ shim nodes are in dark, then it is impossible to 
detect such an attack and replace the primary.
\end{lemma}

\begin{proof}
Let $D$ be the set of shim nodes in dark, such that $\abs{D} \le \f{\Replicas}$. 
We start with the assumption that the nodes in $D$ are able to prove that they 
are under an attack by the byzantine primary $\Primary{}$ and ensure $\Primary$'s replacement 
by convincing a majority of nodes to participate in the view-change protocol.

For a view-change to take place at least $2\f{\Replicas}+1$ nodes need to support such an event. 
As $\Primary{}$ is clever, it ensures that at least $U \ge \f{\Replicas}+1$ honest 
nodes continuously participate in consensus. 
Clearly, $U > D$, which implies that a majority of honest nodes will not request view-change. 
The remaining $\n{\Replicas} - U - D = \f{\Replicas}$ nodes are byzantine and will support the primary in 
this attack.
Moreover, the nodes in set $U$ cannot distinguish between the nodes in set $D$ and the 
up to $\f{\Replicas{}}$ actual byzantine nodes, 
as the byzantine nodes can always request a view-change in an attempt to derail 
the system progress by replacing an honest primary.
Hence, the view-change request by nodes in $D$ will never be successful.
\end{proof}

{\em Featherweight Checkpoints.}
\label{sss:checkpoints}
To recover from nodes in dark attacks, 
we design a featherweight variant of existing checkpoint protocols~\cite{pbft,zyzzyva}.
Existing \BFT{} protocols require 
nodes to periodically construct and exchange $\Name{Checkpoint}$ messages, 
but these messages are {\em expensive} as they include all the client requests and the 
proof that they are committed ($\Name{Commit}$ messages from $2\f{\Replicas}+1$ distinct nodes) 
since the last checkpoint.
As our shim nodes neither execute client requests nor store any data, 
during our featherweight checkpoint protocol, these nodes only send the 
signed proofs (certificates) for each committed request since last checkpoint.

{\bf \em Remark.} 
The nodes in dark attacks do not make the system unsafe but put it at the mercy of the byzantine nodes, which can stop responding after several consensuses have passed; the system suffers from massive communication during recovery.

\subsection{Verifier Flooding}
\label{ss:verifier-flooding}
As the verifier manages all updates to the data-store, 
it is a desirable target by byzantine components.
Specifically, byzantine components can try the following ways to disrupt the system 
by flooding the verifier with redundant requests.
\begin{enumerate}[nosep,wide,label=(\roman*)]
\item \label{flood:prime} {\em Duplicate Spawning by Primary.}
If the shim's primary node is byzantine, it can spawn more executors than necessary.

\item \label{flood:old-prime} {\em Duplicate Spawning by Non-primary.}
A byzantine non-primary node that was once the primary node of shim has access to old 
certificates and $\Name{Execute}$ messages. 
It can use these messages to spawn new executors at the serverless cloud.

\item \label{flood:executor} {\em Duplicate Messages by Executors.}
A byzantine executor can send duplicate $\Name{Verify}$ messages to the verifier.
\end{enumerate}

Although flooding attacks seem trivial to perform, they have monetary impacts on the 
byzantine components. 
Spawning each serverless executor requires the spawner to pay a fixed amount of money.
As a result, any flooding attack performed by a byzantine component will be {\em self-penalizing}.
For example, in our architecture, each primary is paid a fixed amount per consensus 
by the edge application organization.
Hence, a rational byzantine component will avoid this attack.

Moreover, all of these attacks are trying to flood the verifier with the $\Name{Verify}$ messages.
To mitigate the impact of these flooding attacks:
we require the verifier $\Verifier{}$ to ignore any $\Name{Verify}$ 
message for a client request $m$, once it has received matching $\Name{Verify}$ 
messages for $m$ from $\f{\Executors{}}+1$ executors.
Finally, it is a common practice to connect different entities on the network via sockets. 
If flooding attacks take place, the verifier can block communication from such connections.

\section{Transactional Conflicts}
\label{s:conflicts}
Two client transactions $\Transaction{}$ and $\Transaction'$ are 
termed as {\em conflicting}
if $\Transaction{}$ and $\Transaction'$ require access to a common data-item $x$ and at least one of these operations writes to $x$~\cite{distdb}. 
In our \Serverless{} protocol, transactional conflicts arise
from the following set of transactions: 
two transactions $\Transaction{}$ and $\Transaction'$ 
ordered at sequences $k$ and $k'$, respectively, 
and $k < k'$ and $\Transaction$ writes to $x$, which 
$\Transaction'$ reads.

\begin{example}\label{ex:conflict}
For ensuing discussions, we assume two conflicting 
transactions $\Transaction{}$ and $\Transaction'$. 
Let the sequence number for $\Transaction{}$ be $3$ and 
sequence number for $\Transaction'$ be $4$. 
Further, assume $\Transaction$ needs to write to data-item $x$ and 
$\Transaction'$ needs to read $x$.
\end{example}

\subsection{Concurrent Spawning}
\label{s:ooo-spawn}
On a close inspection of Figure~\ref{alg:serverless}, one can observe that the primary 
$\Primary{}$ does not wait for consensus of the $k$-th request to finish before initiating 
consensus for the $(k+1)$-th request.
This process of concurrently invoking multiple consensuses has been employed by prior works  
to increase the system throughput as it reduces the idle times for nodes~\cite{geobft,kauri}.

To further boost the throughput, we permit the primary 
to spawn the $\n{\Executors{}}$ executors for the $(k+1)$-th request prior to spawning executors for 
$k$-th request. 
We term this as {\em concurrent spawning}. 
If the client requests are non-conflicting, concurrent spawning helps to parallelize 
execution.

In the case transactions are conflicting, like $\Transaction{}$ and $\Transaction'$
of Example~\ref{ex:conflict}, we can have two cases:
the read-write sets a transaction accesses are either {\em known or unknown} to the shim 
nodes prior to execution.
Depending on the knowledge of read-write sets, 
transactions {\em may or may not abort} in our architecture.
A naive way would be to ask the shim primary to sequentially spawn executors for each client request, 
but that will significantly reduce the throughput attained by our \Serverless{} protocol.
Hence, we design algorithms to handle either cases, which we discuss next.

\subsection{Unknown Read-Write Sets}
If the shim nodes cannot determine the read-write sets of a transaction during consensus, 
we require the shim nodes to continue following the algorithm in Figure~\ref{alg:serverless}.
The only change is that the shim's primary should spawn an {\em additional} $\f{\Executors}$ executors; 
the shim primary now spawns $\n{\Executors} \ge 3\f{\Executors}+1$ executors instead of $\n{\Executors} \ge 2\f{\Executors}+1$ 
as stated earlier.
We prove the need for these additional executors later.

However, due to the conflicting transactions like $\Transaction{}$ and $\Transaction'$ 
of Example~\ref{ex:conflict}, the verifier $\Verifier{}$ may observe the following:
(i) it did not receive $\f{\Executors}+1$ matching $\Name{Verify}$ 
messages for $\Transaction'$, or
(ii) the read sets of $\Transaction'$ are stale.
In such cases, the verifier would have to abort transaction $\Transaction'$.

{\em Byzantine Aborts and Decentralized Spawning.}
A big challenge to permitting the verifier to abort transactions is 
a byzantine primary that can intentionally delay spawning executors for 
some of the committed transactions to get them aborted.
Moreover, this attack is impossible to detect by other shim nodes or the verifier.
Prior works have shown that there  are no easy solutions to prevent 
byzantine aborts for conflicting transactions with unknown read-write sets~\cite{basil}.
One way to prevent this attack in our serverless-edge architecture is to require each node 
of the shim to spawn some executors at the serverless cloud. 
In specific, once a node $\Replica{} \in \Replicas{}$ commits a client request $m$, it spawns 
$\e$ executors.%
\begin{equation}\label{eq:num-exe}%
\e  = 
\begin{cases}
~1, & \textrm{if } \n{\Executors{}} \le \n{\Replicas}\\
~\left\lceil \dfrac{\n{\Executors{}}}{2\f{\Replicas{}}+1} \right\rceil, & \textrm{otherwise}
\end{cases}
\end{equation}%
If $\n{\Executors{}}$ is less than $\n{\Replicas{}}$, then each node $\Replica{} \in \Replicas{}$
needs to spawn only one executor. 
This will guarantee that of all the spawned executors at least $\f{\Executors{}}+1$ are honest.
Otherwise, each node $\Replica{}$ needs to spawn $\left\lceil \dfrac{\n{\Executors{}}}{2\f{\Replicas{}}+1} \right\rceil$ executors. 
{\em Why?} 
Because up to $\f{\Replicas{}}$ nodes are byzantine and may avoid spawning any executors. 
Hence, the remaining $2\f{\Replicas{}}+1$ honest nodes need to spawn $\n{\Executors{}}$ executors.
Clearly, the total number of spawned executors ($\e \times \n{\Replicas}$) is much larger than the 
required number of executors $\n{\Executors}$.
This is a trade-off we need to pay if we want to {\em decentralize the spawning of serverless executors}.
Another major trade-off of this decentralized spawning is that if the read-write sets are known, 
then each node needs to sequentially spawn executors. 
Hence, like primary (refer to Section~\ref{ss:deterministic-transactions}), each node has to track the dependencies.
Moreover, the proposed value of $\e$ is only valid if each honest node commits the client request. 
If up to $\f{\Replicas{}}$ honest nodes are in dark, then $\e$ changes as follows:
\begin{equation}\label{eq:num-exe2}%
\e  = 
\begin{cases}
~1, & \textrm{if } \n{\Executors{}} \le \n{\Replicas}\\
~\left\lceil \dfrac{\n{\Executors{}}}{\f{\Replicas{}}+1} \right\rceil, & \textrm{otherwise}
\end{cases}
\end{equation}%
Conservatively, we can set $\e = \n{\Executors{}}$, but that will lead to spawning $\n{\Executors} \times \n{\Replicas}$
executors in the worst case.

{\em Verifier Abort Detection.}
\label{ss:verifier-abort}
With the addition of byzantine aborts, 
the verifier needs to determine when to abort a transaction $\Transaction'$ and 
if possible, the cause for abort. 
As a result, the verifier needs to wait for $\f{\Executors{}}+1$ matching $\Name{Verify}$ messages for $\Transaction'$.
For this purpose, our \Serverless{} protocol requires the verifier $\Verifier{}$ to start a timer $\Timer{m}$ 
when it receives the first $\Name{Verify}$ message for the transaction $m := \Transaction'$. 
$\Verifier{}$ stops $\Timer{m}$ when it receives 
$\f{\Executors}+1$ matching $\Name{Verify}$ messages, or 
it receives $\Name{Verify}$ messages from all the $3\f{\Executors}+1$ executors.

Like in Figure~\ref{alg:serverless},
say the verifier collects all the incoming $\Name{Verify}$ messages for $m$ in a set $\Vset$.
If the verifier's timer expires while waiting, it takes one of the following actions:
 
\begin{itemize}[wide]
\item {$\abs{\Vset} < 2\f{\Executors}+1$ }: 
This case implies that the verifier $\Verifier{}$ received less than $2\f{\Executors}+1$ 
$\Name{Verify}$ messages for transaction $\Transaction'$.
As a result, $\Verifier{}$ concludes that the primary $\Primary{}$ is byzantine and it 
creates and broadcasts a $\Name{Replace}$ message to the shim nodes.
Receiving less than $2\f{\Executors}+1$ $\Name{Verify}$ messages implies that either the primary $\Primary{}$ 
spawned less than $\n{\Executors}$ executors or some messages got dropped;
at most $\f{\Executors{}}$ executors can act byzantine and can decide to not send $\Name{Verify}$ messages to $\Verifier{}$, 
In either case, it is safe to conservatively blame the primary. 
Note: even existing \BFT{} protocols decide to blame the primary if messages get dropped~\cite{pbft,sbft,zyzzyva}.

\item {$\n{\Executors} > \abs{\Vset} \ge 2\f{\Executors}+1$ }: 
This case implies that the verifier $\Verifier{}$ received more than $2\f{\Executors}+1$
$\Name{Verify}$ messages for transaction $\Transaction'$.
As a result, the verifier $\Verifier{}$ cannot conclude that the shim's primary is byzantine 
as $\Verifier{}$ has received $\Name{Verify}$ message from 
at least $2\f{\Executors{}}+1$ distinct executors. 
Observing responses from at least $2\f{\Executors}+1$
executors is a guarantee that at least $\f{\Executors}+1$ honest executors 
tried to execute $\Transaction'$ to the best of their ability.
Hence, even if the shim's primary is byzantine and intentionally delays spawning executors for $\Transaction'$, 
there is no way that the verifier can prove this (due to concurrent spawning).

This forces the verifier to abort this transaction.
Assume $k$ is the sequence number for $\Transaction'$.
If $\kmax = k$, then $\Verifier{}$ sends the client an $\SignMessage{\Message{\Name{Abort}}{\Transaction'}}{\Verifier{}}$ 
message.
Otherwise, $\Verifier{}$ adds $\Transaction'$ to the list $\pim$, but {\bf \em tags} it as abort. 
Later, when $\Transaction'$ is extracted from the list $\pim$, the verifier $\Verifier{}$ aborts it.

\end{itemize}

We now describe the indistinguishable attack, 
which forces us to require primary to spawn $\n{\Executors} \ge 3\f{\Executors}+1$.

\begin{theorem}
If client transactions are conflicting and  the primary $\Primary{}$ spawns  
$\n{\Executors} < 3\f{\Executors}+1$ executors, then the \Serverless{} protocol 
faces an indistinguishable attack.
\end{theorem}

\begin{proof}
Assume that $\Primary{}$ spawns only $2\f{\Executors}+1$ executors. 
We know that up to $\f{\Executors{}}$ of these executors can act byzantine. 
As a result, for any client request, the verifier $\Verifier{}$ may receive only 
$\f{\Executors}+1$ $\Name{Verify}$ messages.
Further, due to transactional conflicts, these $\f{\Executors}+1$ $\Name{Verify}$ messages 
may not match.
Eventually, $\Verifier$'s timer will expire and it needs to take some action. 
$\Verifier{}$ can decide to abort this transaction, but this would 
lead to a new problem---a byzantine primary $\Primary{}$ may never 
spawn more than $\f{\Executors}+1$ executors and up to $\f{\Executors{}}$ 
of those executors may be byzantine.
Hence, all subsequent conflicting transactions may abort.

Alternatively, $\Verifier$ can blame the primary 
for receiving less than $\f{\Executors}+1$ matching $\Name{Verify}$ messages, 
but such a decision could be wrong as 
$\Primary{}$ may not be byzantine and 
the lack of sufficient matching messages 
could be a result of conflicts and byzantine executors.%
\end{proof}

\subsection{Best Effort Conflict Avoidance}
\label{ss:deterministic-transactions}
In database literature, several works have employed the concept
of deterministic databases for efficient conflict resolution~\cite{calvin,deneva,qstore}.
In these databases, the 
order in which transactions are applied to the database is determined prior to its execution, which
is only possible if the read-write sets of the transactions are known 
to the participating nodes.

In our \Serverless{} protocol, we learn from these databases. 
If the primary has any knowledge of the read-write sets, 
it uses the {\em queuing strategy} of these databases, 
to create plans that allow running non-conflicting 
transactions in parallel~\cite{calvin,deneva,quecc,qstore,eve}.
Such a strategy would require us to make straightforward modifications to the 
algorithm presented in Figure~\ref{alg:serverless}. 
We would need the shim primary to maintain a {\em logical map} 
of all data-items. 
This map does not store any values of the data-items, but helps the 
primary to {\em locally lock} different data-items.
Further, the primary can no longer concurrently spawn executors for a transaction
until it has determined its conflicts.
Next, we list the steps.

\begin{enumerate}[nosep,wide,ref=\arabic*]
\item The primary $\Primary{}$ adds the $k'$-th transaction to the execution queue after it has 
added (or spawned executors for) all the $k$-th transactions in the queue, where $k < k'$.

\item \label{conf:spawn} If the $k'$-th transaction does not conflict with any $k$-th 
transaction ($k < k'$), $\Primary{}$ spawns serverless executors for the $k'$-th transaction
after it has logically locked all the data-items that are {\em written} by the
$k'$-th transaction.

\item \label{conf:dequeue} Next, $\Primary{}$ dequeues a non-conflicting transaction 
at the head of some queue and repeats Step~\ref{conf:spawn}.

\item When $\Primary{}$ is notified by the verifier $\Verifier{}$ that $\Transaction{}$ 
has been executed, it unlocks the data-items accessed by $\Transaction$  and follows Step~\ref{conf:dequeue}.
We believe these steps can help to reduce aborts.
\end{enumerate}

\section{Safety and Liveness Guarantees}
We now prove that \Serverless{} guarantees safety and 
liveness.
As the shim nodes employ \PBFT{} protocol, we borrow the following 
proposition guaranteed by \PBFT{}.

\begin{proposition}\label{prop:non_divergent}
Let $\Replica{i}$, $i \in \{1, 2\}$,  be two honest shim nodes that committed $\SignMessage{\Transaction_i}{\Client{i}}$ as the $k$-th transaction of view $v$. 
If $\n{\Replicas{}} > 3\f{\Replicas{}}$, then $\SignMessage{\Transaction_1}{\Client{1}} = \SignMessage{\Transaction_2}{\Client{2}}$.
\end{proposition}

\begin{theorem}
Given an architecture $\Arch = \{\Clients, \Replicas, \Executors, \Storage, \Verifier \}$, 
if the number of byzantine shim nodes and byzantine serverless executors are bounded by $\f{\Replicas{}}$ and 
$\f{\Executors}$, respectively, then \Serverless{} protocol guarantees safety.
\end{theorem}

\begin{proof}
Prior to proving this, we note that as the verifier $\Verifier{}$ is
trusted, storage $\Storage$ will be updated in the order 
agreed by $2\f{\Replicas}+1$ of shim nodes. 
We prove the rest as follows:

{\em Non-conflicting transactions.} 
If the primary $\Primary{}$ is honest, then 
from Proposition~\ref{prop:non_divergent}, we can conclude
that no two shim nodes will commit different transactions at the same sequence number and 
$\Primary{}$ will spawn $2\f{\Executors}+1$ executors.
These transactions will persist across views
as in any view-change quorum of $2\f{\Replicas}+1$ replicas, there will be one honest replica that 
has executed this request.
If $\Primary{}$ is byzantine and assigns two or more requests the same sequence 
number $k$, then from Proposition~\ref{prop:non_divergent}, we 
know that $\Primary{}$ will not be successful.
If the byzantine $\Primary{}$ sends the $\Name{Preprepare}$ for some $\Transaction{}$
to less than $2\f{\Replicas}+1$ replicas, this transaction will not commit. 
As a result, at least $\f{\Replicas}+1$ replicas will timeout and a $\Name{ViewChange}$ will take place.
The new primary waits for $\Name{ViewChange}$ messages from $2\f{\Replicas}+1$ replicas, 
and uses these messages to create a $\Name{NewView}$ message.
This $\Name{NewView}$ message includes a list of requests for each sequence number present in the $\Name{ViewChange}$ message.
Each replica on receiving the $\Name{NewView}$ message can verify its contents and 
update its state.

{\em Conflicting transactions with unknown read-write sets.}
In the case of conflicting transactions, the only additional attack a byzantine primary $\Primary{}$ 
can do is to get a transaction aborted by delaying spawning executors.
However, as $\Primary{}$ does not know, which transactions are conflicting, 
this is all based on a guess.
Note: this attack does not make the data-store unsafe.
\end{proof}

\begin{theorem}
Given an architecture $\Arch = \{\Clients, \Replicas, \Executors, \Storage, \Verifier \}$, 
if the network is reliable and 
the number of byzantine shim nodes and serverless executors are bounded by $\f{\Replicas{}}$ and 
$\f{\Executors}$, respectively, then \Serverless{} guarantees liveness.
\end{theorem}

\begin{proof}
Prior to proving this, we note that as the verifier $\Verifier{}$ is
trusted, so if it receives $2\f{\Executors}+1$ matching $\Name{Verify}$ messages 
with correct read-write sets, it will send a reply to the client.
We prove the rest as follows:

{\em Non-conflicting transactions.} 
If the primary $\Primary{}$ is honest, 
then every transaction will be committed by at least $2\f{\Replicas{}}+1$ shim nodes.
$\Primary$ will use this to create a certificate and spawn $2\f{\Executors}+1$ executors and 
$\Verifier{}$ will receive $\f{\Executors}+1$ matching responses.

If $\Primary{}$ is byzantine, it can perform one of the many types of request suspension attacks 
described in Section~\ref{ss:req-suspension}.
For each such attack, either the client $\Client{}$ or the nodes in $\Replicas$ will timeout.
This will force $\Primary{}$ to either ensure consensus of $\Client{}$'s transaction, or be replaced through 
the view-change protocol.
Post view-change, if the subsequent primary is also byzantine, then it will also be eventually replaced.
This process can happen at most $\f{\Replicas{}}$ consecutive 
times, after which the system will be live.
In the case a byzantine primary $\Primary{}$ attempts to keep up to $\f{\Replicas{}}$ nodes in dark, 
then using the featherweight checkpoint protocol these nodes will be brought to the same state.

{\em Conflicting transactions with unknown read-write sets.}
In the case of conflicting transactions, the only additional attack a byzantine primary $\Primary$ 
can do is to spawn less than $3\f{\Executors}+1$. 
In such a case, if the verifier $\Verifier{}$ receives less than $2\f{\Executors}+1$ $\Name{Verify}$ 
messages, such that less than $\f{\Executors}+1$ are matching, 
$\Verifier$'s timer $\Timer{m}$ will timeout and it will send a $\Name{Replace}$ message to the shim nodes.
For other cases, $\Verifier{}$ will send the client a $\Name{Response}$ or $\Name{Abort}$ message 
depending on if it receives  $\f{\Executors}+1$ matching $\Name{Verify}$ messages.
\end{proof}

\section{Implementation}
\label{s:impl}
To gauge the practicality of our vision of a \BFT{} serverless-edge architecture, we implement and evaluate 
our design.

{\em Shim.} As the shim nodes represent edge devices, which may have access to limited 
resources, we want the shim nodes to have a lightweight \BFT{} implementation.
So, on each shim node, we install \ResilientDB's node architecture~\cite{geobft,rcc,r-evalpaper,poe,resilientdb-demo,tut-vldb20}. 
\ResilientDB{} provides access to a multi-threaded, pipelined, and modular architecture  
for designing \BFT{} applications.%
\footnote{
\ResilientDB{} is open-sourced at {\em https://resilientdb.com/}.
}
The codebase is written in C++ and we deploy \ResilientDB's \PBFT{} protocol at the shim.
Clients also employ C++ to create YCSB transactions (refer Section~\ref{s:eval}) and use NNG~\cite{nng} sockets for communication.

{\em Invoker.} 
At each shim node, we deploy an invoker to spawn $\n{\Executors}$ executors 
when indicated by the node's consensus instance.
\ResilientDB{} provides at each node an {\em execute-thread}, which  
calls invoker as soon as a request is committed.
Our implementation of the invoker is written in Go~\cite{go} using the AWS SDK for Go.
Further, our invoker does not wait for the spawned executors to finish and proceeds to spawn the executors for the next 
client request.

{\em Serverless Function.} 
Each AWS Lambda executor receives a function written in C++ that includes the client transaction. 
This function instructs the executor to: (i) verify the certificate $\mathfrak{C}$, 
(ii) execute the transaction, 
(iii) fetch necessary read-write sets from the storage database, and 
(iv) send the result to the verifier.
We encode the communication between the Lambda function and the verifier in a stateless HTTP request. 
We use CryptoPP\cite{cryptopp} library for digital signatures and verification and 
use CPR\cite{cpr} to create and send HTTP requests.

{\em Verifier.}  
We implement the verifier in Go and install a simple HTTP/Net webserver at the verifier for receiving the executor responses.
Further, our verifier includes a {\em hashmap} to count the matching responses for each transaction.
Post validation, the verifier uses NNG to send a response to the client.

\section{Evaluation}
\label{s:eval}
Our evaluation aims to answer following questions regarding our \Serverless{} protocol.
\begin{enumerate}[wide,nosep,label=(Q\arabic*),ref={Q\arabic*}]
    \item \label{Q:lat-tput} Impact of client congestion?
    \item \label{Q:scale-executors} Impact of increasing the number of executors?
    \item \label{Q:batch-size} Impact of batching client requests?
    \item \label{Q:length-execution} Impact of expensive execution?
    \item \label{Q:exe-placement} Impact of spawning executors across globe?
    \item \label{Q:cores} Impact of resource limitations at edge devices?
    \item \label{Q:conflict} Impact of conflicting transactions?
    \item \label{Q:perf} Baseline comparison of \Serverless{}?
    \revise{
    \item \label{Q:offload} Impact of task offloading?
    }
\end{enumerate}

{\bf \em Setup.}
We deploy the verifier, shim nodes, and clients on the Oracle Cloud Infrastructure (OCI). 
These components use {\em VM.Standard.E3.Flex} architecture with \SI{10}{GiB} NICs.
Each shim node has $16$ cores and \SI{16}{GiB} RAM and
the verifier has $8$ cores.
We use AWS Lambda Functions for spawning serverless executors in up to $11$ regions in the following order: 
North California, Oregon, Ohio, Canada, Frankfurt, Ireland, London, Paris, Stockholm, Seoul, and Singapore.
In our experiments, we use up to $128$ shim nodes and $21$ executors.
We run each experiment for $180$ seconds with $60$ seconds warmup time and report
the average results over three runs.

Unless {\em explicitly} stated, we use the following setup.
We require the primary node to spawn $3$ AWS Lambda executors,
each of which is spawned in a distinct region. 
Further, we deploy up to \SI{80}{k} clients on $4$ OCI machines to concurrently issue requests.
Each client waits for a response prior to sending its next request.
We also require clients and edge nodes to employ {\em batching} and 
run consensuses on batches of $100$ client transactions.
The size of each type of message communicated is:
$\Name{Preprepare}$  (\SI{5392}{B}),
$\Name{Prepare}$ (\SI{216}{B}),
$\Name{Commit}$ (\SI{220}{B}),
$\Name{Execute}$ (\SI{3320}{B}), and
$\Name{Response}$ (\SI{2270}{B}).

\revise{
{\bf \em Benchmark.}
To evaluate our serverless-edge architecture across different parameters, 
for some experiments, we need to fix the number of shim nodes. 
We learn from existing database literature, specifically the Blockbench~\cite{blockbench} paper, 
and select two configurations.
\ServEB{}: Medium size shim with $8$ nodes.
\ServTB{}: Large size shim with $32$ nodes (maximum number of nodes in any Blockbench experiment).

Similarly, we adopt the popular {\em Yahoo Cloud Serving Benchmark} (YCSB) 
from Blockbench suite, 
which has also been used by several prior works in database literature for designing 
transactions~\cite{geobft,qstore,blockbench,ycsb,easyc}.
}
We use YCSB to create key-value transactions that access a database of 
$\SI{600}{\kilo\nothing}$ records.
Specifically, our transactions perform read and write operations.
With regards to edge applications, these transactions represent user transactions 
that require access to existing records in the storage.

\begin{figure}
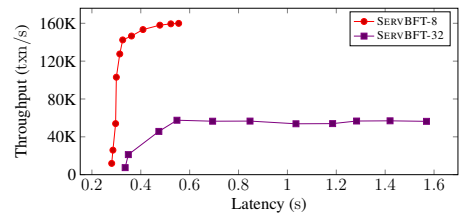

    \centering
    \graphE
    \caption{Comparing latency against throughput on varying the number of clients sending requests to the shim.}
    \label{fig:LatTP}
\end{figure}

\begin{figure*}
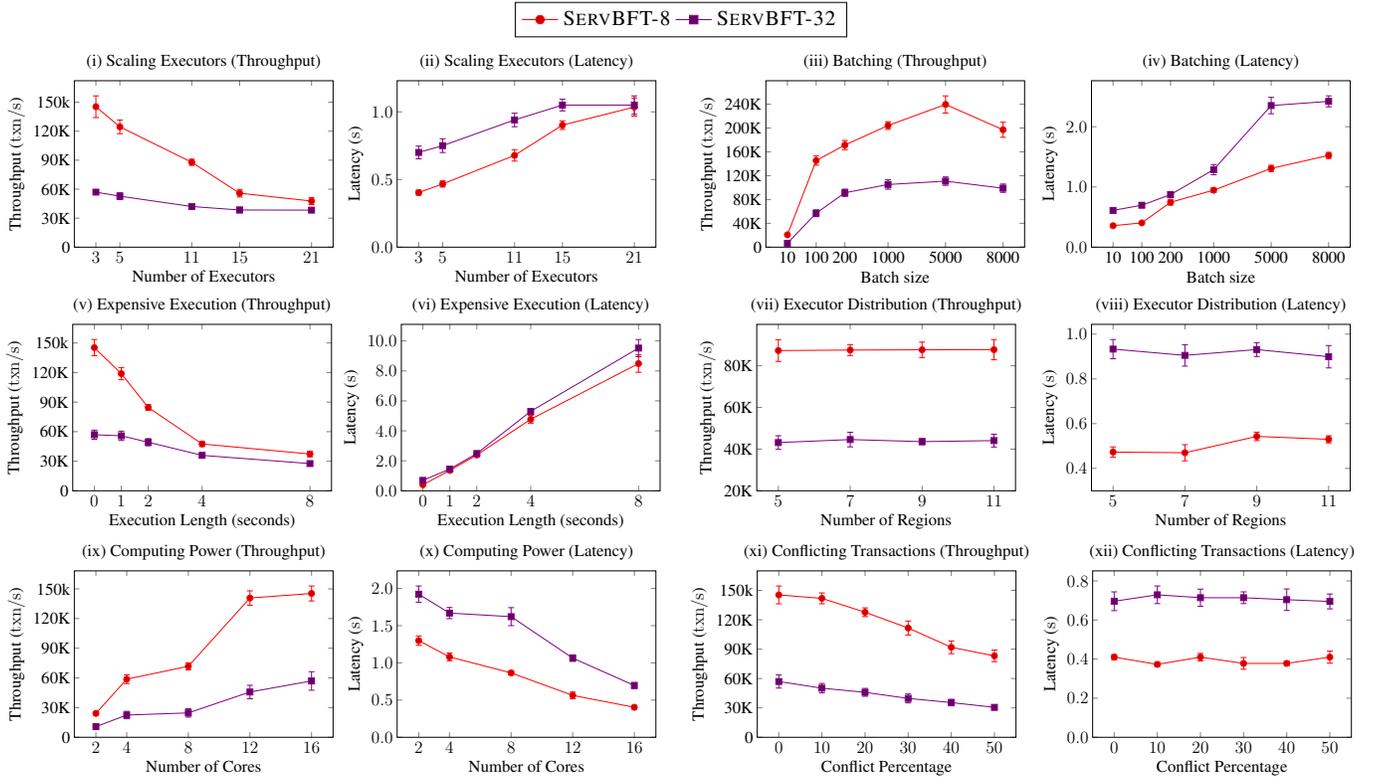

    \centering
    \setlength{\tabcolsep}{1pt}
    \scalebox{0.8}{\ref{mainlegend}}\\[5pt]
    \begin{tabular}{cc@{\quad}cc}

        \graphBTP & \graphBLat &  \graphCTP & \graphCLat \vspace{0.1cm} \\ 
        \graphDTP & \graphDLat &  \graphFTP & \graphFLat \vspace{0.1cm} \\
        \graphGTP & \graphGLat &  \graphHTP & \graphHLat
    \end{tabular}
    \caption{Benchmarking throughput attained and latency incurred by the \BFT{} Serverless-Edge architecture.}
    \label{fig:plots}
\end{figure*}

%\begin{enumerate}[wide,label={\bf \em \Alph*.},ref={\ref{s:eval}-\arabic*}]
%\item {\bf \em Impact of Client Congestion.} 
\subsection{Impact of Client Congestion}
\label{ss:client}
In Figure~\ref{fig:LatTP}, we vary the number of deployed clients from \SI{2}{k} to \SI{88}{k}. 
For the first five data-points on the graph, we double the number of clients and for succeeding points, 
we increase the number of clients by \SI{8}{k}.
Initially, an increase in the number of clients causes an increase in system throughput, post which
the throughput saturates.
This happens because {\em each entity in our serverless-edge architecture has to now do more work than before, 
which causes an increase in computational and communication costs.}
As a result, the latency keeps increasing as each request spends a longer time in the architecture.
Hence, \ServE{} outperforms \ServT{} as {\em fewer nodes are involved in each consensus, which implies 
smaller wait time for each request.} 
{\em Summary:} 
We observe that initially \ServE{} attains up to $1.6\times$ more throughput and 
$1.2\times$ less latency than \ServT{}.
However, on increasing the number of clients, the gap increases to 
$2.8\times$ more throughput and  $2.71\times$ less latency.

%\item {\bf \em Impact of Executors.}
\subsection{Impact of Executors}
\label{ss:executors}
In Figures~\ref{fig:plots}(i) and~\ref{fig:plots}(ii), we vary 
the number of serverless executors spawned by the primary node: $3$, $5$, $11$, $15$, and $21$. 
For these experiments, we spawned executors in up to {\em seven} regions and tried to 
evenly split these executors across these regions.
These figures illustrate that an increase in the number of executors causes a decrease in throughput and 
an increase in latency.
Although all the executors process the requests in parallel, 
{\em there is an increase in the task of spawning at the primary and increase in validation at the verifier.}
Further, as executors are spread across distinct regions, {\em the reduced bandwidth and increased ping costs}
delays communication.
{\em Summary:} At $3$ executors, \ServE{} attains $2.59\times$ more throughput and $43\%$ less latency than \ServT{}, 
while at $15$ executors, $47\%$ more throughput and $5\%$ less latency.

%\item {\bf \em Impact of Batching.}
\subsection{Impact of Batching}
\label{ss:batching}
In Figures~\ref{fig:plots}(iii) and~\ref{fig:plots}(iv), we vary the size of batch of client requests from $10$ to 
\SI{8}{k}.
With an increase in batch size, we first observe an increase in the system throughput followed by an eventual decrease.
Although larger batches imply a corresponding decrease in the number of runs of the \Serverless{} protocol, 
{\em it substantially increases the costs of communicating batches across the shim nodes and executors.}
Further, larger batches are much more expensive to process for shim nodes and executors.

{\em Summary:} From batch size $10$ to \SI{5}{k}, \ServE{} observes an increase in throughput by $11.42\times$ 
and \ServT{} observes an increase in throughput by $18.5\times$.

%\item {\bf \em Impact of Expensive Execution.}
\subsection{Impact of Expensive Execution}
\label{ss:expensive}
In Figures~\ref{fig:plots}(v) and~\ref{fig:plots}(vi), we test with transactions that require large 
execution time;
we vary the time required for execution from few milliseconds to $8$ seconds.
As the time required to execute a transaction increases, the time required by the 
shim and the verifier to {\em process this request becomes insignificant.} 
Prior works show that such transactions or codes, which 
bottleneck the system throughput and latency  are prevalent~\cite{hyperledger-fabric}.
This experiment also proves that {\em our serverless-edge architecture introduces minimal costs} to
the applications that require large execution times.
{\em Summary:} From execution length of few milliseconds to $8$ seconds, \ServE's 
throughput reduces by $74.5\%$ and latency increases by $21\times$, while \ServT's throughput reduces 
by $51\%$ and latency increases by $13.6\times$.

%\item {\bf \em Impact of Spawning Executors across Globe.}
\subsection{Impact of Spawning Executors across Globe}
\label{ss:executor-regions}
In Figures~\ref{fig:plots}(vii) and (viii), we require the primary node to spawn $11$ 
executors in $5$, $7$, $9$, and $11$ regions; 
we vary the number of regions while spawning same number of executors.
The primary node uses the round-robin protocol to spawn executors in each region.
In this experiment, we want to observe the impact of system performance on increasing the number of regions. 
We observe that the {\em throughput and latency remain constant.} 
The primary node spawns $11$ executors ($\f{\Executors} = 5$), so the verifier needs to wait for 
only $\f{\Executors}+1 = 6$ matching $\Name{Verify}$ messages. 
The first $6$ messages received by the verifier (deployed at North California) 
are from nearby regions: North American and European.

%\item {\bf \em Impact of Computing Power.}
\subsection{Impact of Computing Power}
\label{ss:cores}
We use Figures~\ref{fig:plots}(ix) and~\ref{fig:plots}(x) to limit the available 
computing resources at shim nodes.
As shim nodes represent edge devices, these devices  
may have limited cores and memory. 
So we test the impact of this restricted hardware om \Serverless{}.
Unsurprisingly, as we increase the number of available cores, the protocols 
achieve higher throughputs and lower latencies.
This is the case because {\em our shim nodes adopt the multi-threaded pipelined 
architecture} of \ResilientDB{}, which performs better with an increase in 
available cores.

{\em Summary:} From experiments at $2$ cores to $16$ cores, \ServE's 
throughput increases by $6\times$ and latency decreases by $70\%$, while \ServT's throughput increases 
by $5\times$ and latency decreases by $64\%$.

%\item {\bf \em Impact of Conflicting Transactions.}
\subsection{Impact of Conflicting Transactions}
We now vary the degree of transactional conflicts from $0$ to $50\%$
and illustrate our findings in Figures~\ref{fig:plots}(xi) and~\ref{fig:plots}(xii).
{\em As the read-write sets are unknown, the primary node cannot lock the conflicting 
data-items.}
As a result, some transactions get aborted at the verifier.
This leads to an expected observation--a decrease in throughput with an increase in the rate of conflicts.
However, the latency remains unchanged as the response time for the client remains the same.

{\em Summary:} From $0\%$ conflicting transactions to $50\%$ conflicting, \ServE's 
throughput decreases by $43\%$, and \ServT's throughput decreases by $46\%$.

\begin{figure}
    \centering
    \setlength{\tabcolsep}{1pt}
    \scalebox{0.8}{\ref{mainlegend2}}\\[5pt]
    \begin{tabular}{cc}
        \graphATP & \graphALat
    \end{tabular}
    \caption{Comparing \Serverless{} against our three baseline designs: \ServerlessCFT{}, \PBFT{} and \NoShim.}
    \label{fig:graph-comp}
\end{figure}

%\item {\bf \em Shim Scalability.}
\subsection{Shim Scalability}
\label{ss:shim}
Until now, in all the experiments, we ran the \PBFT{} protocol at the shim. 
So, we create {\em three baseline designs} to compare against \Serverless{}:
\begin{enumerate}[nosep,wide,label=(\alph*)]
\item \NoShimB -- Represents the experiment where there is no shim; no \BFT{} consensus takes place. 
All the clients send their requests to a node, which instantaneously spawns executors.

\item \ServerlessCFTB -- Represents the experiment where the shim nodes employ a crash fault-tolerant 
(\CFT{}) like Paxos~\cite{paxos} for consensus. 
As \CFT{} protocols do not protect against byzantine attacks, they do not require cryptographic 
signatures, which in turn reduces the amount of work done per consensus. 
Further, unlike \PBFT{}, Paxos is linear.

\item \PBFTB -- We also test our \Serverless{} protocol against a \BFT{} system 
(e.g. \ResilientDB{})
running the \PBFT{} protocol. 
In this system, we assume each node is a replica and executes the request in the agreed order post consensus~\cite{pbft,geobft}. 
As a result, there are no costs associated with spawning executors and waiting for verifier to validate 
the requests.
\end{enumerate}

In these experiments, we also gauge how the shim scales with an increase in the number of edge devices. 
For this purpose, we vary the number of shim nodes from $4$ to $128$.
We use Figure~\ref{fig:graph-comp} to illustrate the throughput and latency metrics and
observe the following order for throughput attained:
\[
\Serverless{} < \PBFT{} < \ServerlessCFT < \NoShim 
\]
\NoShim{} has a constant throughput because there is no change in the number of shim nodes.
Moreover, \PBFT{} performs slightly better than our \Serverless{} protocol. 
This implies that the {\em verifier and executors do not adversely impact the throughput of \PBFT{}.}
Finally, \ServerlessCFT{} outperforms \PBFT{}, which implies that {\em the throughput of the serverless-edge 
architecture can be increased by replacing \PBFT{} with faster consensus protocols.}
{\em Summary:} \Serverless{} and \ServerlessCFT{} achieve up to $22\%$ less throughput   
and $1.25\times$ more throughput than \PBFT{}, respectively.

\begin{figure}
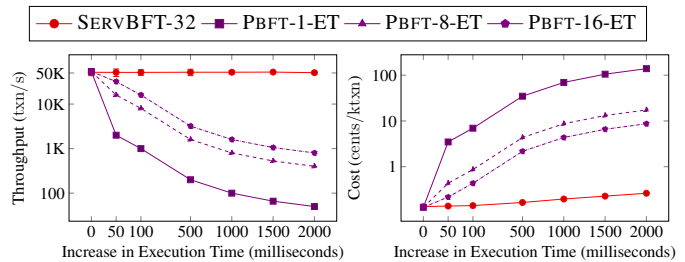

    % \vspace{-5mm}
    \centering
    \scalebox{0.8}{\ref{mainlegend3}}\\[5pt]
    \begin{minipage}{0.49\columnwidth}
        \graphITP 
    \end{minipage}
    \begin{minipage}{0.49\columnwidth}
	\graphJ
    \end{minipage}
    %\vspace{-2mm}
    \caption{\revise{
	Comparing our serverless-edge model against \PBFT{}. 
	Here, ET refers to number of execution threads assigned to specific \PBFT{} implementation.
    }}
    \label{fig:expensive}
    % \vspace{-5mm}
\end{figure}

%\item {\bf \em Impact of Task Offloading.}
\subsection{Impact of Task Offloading}
We use Figure~\ref{fig:expensive} to illustrate the benefits of employing our serverless-edge model. 
Specifically, we introduce compute-intensive tasks (increasing execution time) and compare
the peak throughput and monetary costs against setups where
all the computations (\PBFT{} consensus and transaction execution) are done on the edge devices 
(no serverless).
For this experiment, we compare our \ServT{} with $3$ serverless executors against a \PBFT{} shim with  $32$ nodes.

We make two observations:
(1) If transactions can be executed in parallel, our serverless-edge model is only bounded 
by the rate of consensus and the number of executors that can be spawned in parallel.
This is in contrast to setups where shim performs all tasks and becomes resource-bounded, which
adversely decreases the throughput.
To further validate this resource-boundedness, we calculate monetary costs of these experiments (in $\si{cents\per\k\txn}$)
and use the precise costs for spawning serverless executors at AWS Lambda and running machines on OCI.
Resource-boundedness increases monetary costs as machines need to be run for a larger period of 
time to complete the same set of transactions.
(2) Serverless clouds permit selecting optimal hardware. 
To illustrate this, for experiments where shim does all tasks, we 
vary the number of execution threads (ET) at shim nodes ($1,8,16$). 
If the available hardware has few cores, then a smaller set of transactions ($1$ or $8$) 
can execute in parallel, which impacts throughput.
Alternatively, an enterprise can require edge devices to have more cores ($16$),
which may be underutilized if there is less available parallelism.

%\end{enumerate}

\section{Related Work}
\label{s:related}
{\em Edge computing} is a decade old problem for 
which prior works have presented several interesting 
solutions~\cite{edge-facial-recog,edge-node-placement,edge-challenges,edgelstm,croesus}. 
These solutions aim to reduce latency for edge applications, but they cannot handle byzantine attacks and require 
developers to perform managerial tasks.

In recent years, {\em Serverless computing} has also gained a lot of interest 
with the aim of offloading the managerial tasks such as server provisioning and resource scaling to 
the cloud provider while the developer only uploads the code required to be executed~\cite{serverless-trends,berkeley-serverless}.  
Prior works have presented novel solutions in this direction: 
AFT~\cite{aft} introduces a shim to make stateful executors consistent; and
Faasm~\cite{faasm} aims to design efficient stateful executors.
However, neither these works target edge applications, nor they consider byzantine attacks.

To design applications that can handle byzantine attacks, existing works have employed 
{\em Byzantine Fault-Tolerant} consensus protocols in the context of 
{\em blockchain technology}~\cite{sharper,basil,byz-agreement-comp,blockchaindb,atomic-commit-blockchain,seemore,fides,bft-review,narwhal-tusk,hyperledger-study,poc-ieee}.
These applications assume that a set of nodes holding the same data run a \BFT{} protocol. 
Each committed transaction is noted in an append-only ledger, blockchain, which can be queried in future 
to track transactions.
EdgeChain~\cite{edgechain} introduces a blockchain layer in the 
edge-compute model, which allocates the resources to edge devices.
However, it does not tackle byzantine attacks from edge-clouds.
Bajoudah et al~\cite{iot-data-trading} introduce a blockchain-based edge model where 
IOT devices maintain the blockchain.
Blockene~\cite{blockene} wants to allow mobile devices to participate in 
blockchain consensus by delegating all the storage, computation, and communication 
tasks to a set of powerful servers.

\revise{
Aslanpur et al.~\cite{serverless-edge-vision} present 
the vision of a serverless-edge framework. 
Their proposal does not assign tasks to edge devices and delegates all jobs to the serverless cloud.
Further, there is no discussion on handling byzantine failures. 
Moreover, their vision is neither implemented nor does their paper present any evaluation. 
Baresi et al.~\cite{empower-serverless-edge} present a similar design, 
but their design focusses on mobile computing.
They do present a small evaluation of their design, but neither is their code available, 
nor do they make use of actual serverless cloud providers (like AWS).
Their design delegates everything to the mobile edge servers (where they create a serverless cloud)
and does not handle byzantine failures.
Our NoShim experiment (Figure~\ref{fig:graph-comp}) approximates their architecture.

In comparison, our serverless-edge co-design 
handles byzantine attacks,
permits edge devices to select any serverless provider in vicinity, 
offloads compute-intensive tasks to cloud while allowing light-weight 
ordering on edge devices.
}

\section{Conclusions}
\revise{
In this paper, we presented \Serverless{}, the first protocol 
to guarantee Byzantine Fault-Tolerant transactional flow 
among edge devices and serverless functions.
\Serverless{} facilitates collaboration among edge devices, 
which spawn serverless executors at one or more cloud providers 
in their vicinity to process compute-intensive operations.
Our proposed architecture ensures that only consistent updates 
are written to the database.
We also present solutions to resolve various attacks on our
proposed architecture. 
Our extensive evaluation illustrates that our architecture is scalable 
and is a good fit for the emerging edge applications.
}

\section*{Acknowledgments}
We would like to thank anonymous reviewers who helped improve this paper.
This work was supported in part by (1) Oracle Cloud Credits \& Research Program, and (2) the NSF STTR Award \#2112345 through Moka Blox LLC.

\bibliographystyle{IEEEtran}
\bibliography{reference}

\end{document}